\documentclass[12]{article}
\usepackage{graphics}
\usepackage{epsfig}
\newcommand{\beq}{\begin{equation}}
\newcommand{\eeq}{\end{equation}}

\newcommand{\beqs}{\begin{eqnarray}}
\newcommand{\eeqs}{\end{eqnarray}}

\newcommand{\chose}[2]{\left[ \begin{array}{c} #1 \\ #2 \end{array} 
\right]}
\newcommand{\symmat}[4]{\left[ \begin{array}{cc} #1 & #2 \\ #3 & #4 \end{array} \right]}

\newcommand{\ket}[1]{| #1 \rangle}
\newcommand{\poisn}[2]{\left[ #1 , #2 \right]}
\begin{document}
\bibliographystyle{h-physrev}
\input{epsf}

\title{Yangian Symmetries of Matrix Models and Spin Chains: The Dilatation Operator of $\cal N$$=4$ SYM}
\author{A.Agarwal\thanks{abhishek@pas.rochester.edu} \\
\and
 S.G.Rajeev\thanks{rajeev@pas.rochester.edu} \\
University of Rochester. Dept of Physics and Astronomy. \\
Rochester. NY - 14627}
\maketitle

\begin{abstract}
We present an analysis of the Yangian symmetries of various bosonic sectors of the dilatation operator of $\cal N$$=4$ SYM. The analysis is presented from the point of view of Hamiltonian matrix models. In the various $SU(n)$ sectors, we give a modified presentation of the Yangian generators, which are conserved on states of any size. A careful analysis of the Yangian invariance of the full $SO(6)$ sector of the scalars is also presented in this paper. We also study the Yangian invariance beyond first order perturbation theory. Following this, we derive the continuum limits of the various matrix models and reproduce the sigma model actions for fast moving strings reported in \cite{Kruczenski-1, tseytlin-2,hernandez-1,kristjansen-2, tseytlin-1, tseytlin-3}. We motivate the constructions of continuum sigma models (corresponding to both the $SU(n)$ and $SO(n)$ sectors)  as variational approximations to the matrix model Hamiltonians. These sigma models retain the semi-classical counterparts of the original Yangian symmetries of the dilatation operator. The semi-classical Yangian symmetries of the sigma models are worked out in detail. The zero curvature representation of the equations of motion and the construction of the transfer matrix for the $SO(n)$ sigma model obtained as the continuum limit of the one loop bosonic dilatation operator is carried out, and the similar constructions for the $SU(n)$ case are also discussed. 
\end{abstract}

\section{Introduction and Summary:} In this paper, we extend the work presented in \cite{abhi-N=4-1} and  carry out an analysis of the Yangian invariance of the Dilatation operator of superconformal Yang-Mills theory from the point of view of Hamiltonian matrix models. We use and extend the formalism developed in \cite{rajeev-lee-prl, rajeev-lee-review} to carry out this analysis. We focus mainly on the  (the $SU(n)$ and $SO(n))$ bosonic sectors of the  Dilatation generator. In this analysis, special attention is paid to two particular aspects of this hidden symmetry of the gauge theory. On one hand, we construct, in  rather explicit forms, the $\it{conserved}$ non-local charges for the spin chain in the $SU(n)$ sector. Although the study of non-local conserved currents for theories defined on Lattices has a thorough literature devoted to it, the construction of corresponding conserved charges has remained a  delicate issue. Usually, the conservation of the charges is violated by boundary terms. In this paper we find that these problems (in the one-loop $SU(n)$ sectors) have rather simple resolutions, thereby realizing the Yangian as a true symmetry of the Dilatation operator. The Yangian symmetry of the two loop corrected $SU(2)$ invariant dilatation operator is also analyzed in some detail.   The other aspect of our analysis has to do with the contraction of the Yangian invariance in the continuum limits of the matrix models.  We construct the sigma model Hamiltonians arising as the continuum limits of the matrix models both in the $SU(n)$ and $SO(n)$ sectors.  These continuum sigma models are derived as variational approximations to the matrix model Hamiltonians. We pay special attention to the   analysis of  the $SO(n)$ invariant sigma model obtained as a continuum limit of the Minahan-Zarembo spin chain. We use the formalism developed in \cite{Kruczenski-1, tseytlin-2, tseytlin-1, tseytlin-3} to carry out the analysis. By constructing the monodromy matrix following from the zero curvature representation of the sigma model,  we show that the Yangian symmetry of the spin chain contracts to the $\it {semi-classical}$ $SO(n)$ Yangian. This puts the sigma model on the same footing as the classical Heisenberg model and its $SU(n > 2)$ cousins. 

Since the BMN proposal\cite{bmn}, various important steps  have been taken  in the direction of understanding the $AdS-CFT$\cite{adscft} correspondence better. The basic idea put forward in \cite{bmn} was that for operators containing large $R$ charge $J$, one could do perturbative expansions in an effective t'Hooft coupling $\lambda ' = \frac{\lambda}{J^2}$ and match the perturbative gauge theory computations with those from the string theory side. On the gauge theory end, this amounts to looking at operators which are very nearly chiral primaries, while on the string theory side one essentially looks at strings in a near flat plane wave background. Both ends of the correspondence being tractable has led to vigorous investigations in the recent past. For a detailed account of the BMN correspondence we refer the reader to \cite{bmn-rev, bmn-rev-1} and references therein. On the SYM side, testing the proposal necessitates a good understanding of  the resolution of operator mixing in the gauge theory.  A remarkable step in this direction  has been the observation of Minahan and Zarembo that the one loop dilatation operator of $ \cal N$$=4$ SYM (restricted to the bosonic sector) can be interpreted as  the Hamiltonian of an integrable  $SO(6)$ spin chain, which can be understood and integrated using the algebraic Bethe ansatz\cite{minahan-spin}. This implies that the transfer matrix of the spin chain,  can be diagonalized explicitly, and the spectrum of anomalous dimensions can be computed. It should be noted however that, while this can be achieved in principle, in practice, getting the spectrum for chains of any length is quite intricate, and the analysis  simplifies  only  in the limit of large chains. \footnote{One can go beyond computing the spectrum of the dilatation operator and use the underlying Bethe ansatz to compute correlation functions of the gauge theory as well. This has been demonstrated in \cite{roiban-corr-fn}.}If one goes beyond the scalar sector, then one finds that integrability is not lost, and  the complete one-loop dilatation operator, with Fermions and gauge fields included, can be interpreted as the Hamiltonian of  a  $PSU(2,2|4)$ integrable spin chain\cite{beisert-et-al-super-spin, beisert-et-al-complete-one-loop}. 

By now there are various pieces of evidence that seem to indicate that integrability is not an artifact of the one-loop approximation. At higher loops operator mixing gets progressively complicated, but one can, never the less find closed sub sectors of gauge theory operators within which the problem takes on more tractable forms. One such sub-sector whose study has proved to be extremely fruitful consists of two scalars  transforming among each other by $SU(2)$ rotations\cite{beisert-et-al-conformal}.  The two scalars are taken to be charged under different $U(1)$'s of $SO(6)$. At one loop, the dilatation operator restricted to this sector becomes the familiar $XXX$ Heisenberg spin chain. 
\beq
D_2 = 2\frac{\lambda }{16\pi ^2}\left(\sum_l(I_{l,l+1}-P_{l,l+1})\right).
\eeq
$P_{l,n}$ is the permutation operator that interchanges the spins at sites $l$ and $n$, while $I$ is the identity. 
The higher loop dilatation generators for the $SU(2)$ sector are being studied quite extensively in the current literature, see for example \cite{beisert-et-al-long-range, minahan-et-al-su2,  long-range-inz}. The two loop generator has been derived in \cite{beisert-et-al-conformal}, see also \cite{gross-et-al-1, gross-et-al-2} and takes the form of a spin chain with next to nearest neighbor interactions. 
\beq
D_4 = \left(\frac{\lambda }{16\pi ^2}\right)^2\left(\sum_l(-6I_{l,l+1} 
+ 8P_{l,l+1} -2P_{l,l+2})\right).
\eeq
For such spin chains, it is not clear what the analogs of the algebraic Bethe ansatz is. However, higher conserved charges for this two loop Dilatation generator have been  constructed \cite{beisert-et-al-conformal} indicating a degenerate spectrum; a basic requirement for integrability. As one goes to higher orders in perturbation theory, the range of interaction of the spin chain keeps increasing, and one is sure to get a spin chain with long range interactions as the all loop dilatation generator. In the $SU(2)$ sector, higher loop dilatation generators have been proposed, based on symmetry considerations, integrability  and BMN scaling \cite{beisert-et-al-conformal, beisert-et-al-long-range, long-range-inz}, and it is believed that the long range spin chain is of a hyperbolic type , which interpolates between the Haldane-Shastry model (having infinite range of interaction) and the Heisenberg type, which has only short ranged interactions. While the integrability of these long range spin chains is not a completely settled issue, remarkable progress has been achieved in this direction \cite{beisert-et-al-long-range, long-range-inz, inz-rev}. It should be noted however, that integrability, at least a low orders in perturbation theory is not restricted to the case of super-conformal Yang Mills theory. For a review of integrable structures in the context of light front QCD  see  \cite{belitsky-review} and references therein. For more recent results in the direction of the use of the light cone formalism to understand  the dilatation operator of Yang-Mills theories with various degrees of supersymmetry see \cite{belitsky-light-front}.  Interesting developments in this direction in the context of the renormalization of self-dual components of the field strength in non-supersymmetric Yang-mills theory has been achieved in \cite{ferretti-zarembo}. For a discussion of one loop integrability in $\cal{N}$ $=2 SYM$, we refer the reader to \cite{n=2}. Integrable deformations of $\cal N$$=4 $ $SYM$ have been  been analyzed in considerable detail in \cite{david-b}. 

A parallel set of approaches towards understanding the spectrum of anomalous dimensions of superconformal Yang-Mills theory should also be mentioned. The plane wave limit of matrix theory provides an alternative description of the same problem. In this description, integrability to at-least the third order in perturbation theory has been shown to exist in \cite{plefka-matrix-1, plefka-matrix-3, plefka-matrix-2}. On the other hand, an interesting framework for the computation of anomalous dimensions, based on the representations of higher spin symmetry algebras has been proposed in \cite{bianchi-hs}.  One of the interesting features of the methodology used in this way of computing anomalous dimensions is that it does not make any direct reference to the underlying integrable structures present in the gauge theory.  

On the string theory side, the discovery of various integrable structures have largely facilitated the comparison to results obtained in $SYM$ computations. The string sigma model in $AdS_5\times S^5$ exhibits classical integrability. The existence  of Pohlmeyer charges and Yangian symmetry  of  the sigma model is believed to be the  reason behind its classical solvability \cite{hidden-string-symm-2, hidden-string-symm-3, hidden-string-symm-4, hidden-string-symm-1}. For a related development, see also \cite{das-et-al-2}. Integrability of the string sigma model has been utilized beautifully by Frolov and Tseytlin  to carry out semi-classical computations with the sigma model and obtain predictions for $SYM$ computations \cite{frolov-tseytlin-early, frolov-tseytlin-mul-spin}. Their key insight lay in the realization that a BMN like proposal can be made to work  even without taking a particular limit of the $AdS_5\times S^5$ background. One can invent a new parameter, which is basically the length $J$ of the spin chain (or equivalently an angular momentum like observable on the string side),  and quite like the BMN case, the limit of large $J$ is again accessible from both the gauge theory and gravity sides. Various implementations of this idea were carried out in \cite{ext-fs, ext-fs-1, ext-fs-2, ext-fs-3, ext-fs-4, gleb-matt-1, gleb-matt-2} and  review of this approach can be found in \cite{tseytlin-rev}. In this series of investigations comparisons of both the integrable structures, i.e the higher conserved charges, as well as that of the spectrum of the dynamical systems obtained  from both the sides was carried out. Gauge theory duals of semi-classical string solutions have been discussed in depth in  \cite{gt-dual-1, engquist-1, gt-dual-2, gt-dual-3}. The upshot of this course of investigation has been the discovery of precise agreement between the leading order and the next to leading expressions for  string energies and higher conserved charges, (which are in general intricate functions of $\lambda and  J$), and the corresponding one and two loop gauge theory computations.   However, as yet unresolved discrepancies continue to remain at three and higher loops\cite{gleb-matt-1, gleb-matt-2, gt-dual-2, callan-et-al-3}. 

A fresh insight in this program was brought about by the developments initiated by Kruczenski \cite{Kruczenski-1}. It was shown in \cite{Kruczenski-1}, that the effective semiclassical action, for a string rotating with a large angular momentum can be mapped to the one describing the continuum limit of the Heisenberg spin chain, which is nothing but the one loop dilatation operator restricted to the $SU(2)$ sector. This allows one to carry out comparisons at the level of actions, and provides  an explanation for the various agreements found previously. This point of view has since then been pushed further, and  it has been shown, that for various $SU(n)$ sectors the effective action for the sigma model on $AdS_5\times S^5$, describing fast moving strings on the $S^5$, agrees with those arising as the continuum limit of $SU(n)$ spin chains\cite{krc-2,hernandez-1,kristjansen-1,kristjansen-2, tseytlin-1, tseytlin-2, tseytlin-3}. Even for the full $SO(6)$ scalar sector, this program has been shown to be very promising \cite{tseytlin-2, tseytlin-1}. 

A key theme tying together these integrable structures is that of  Yangian symmetries. These Hopf algebraic symmetries are, to a large extent responsible for integrability on both the sides. The fact that Yangian symmetries are present at the level of the string sigma model was elucidated in \cite{hidden-string-symm-2, hidden-string-symm-3, hidden-string-symm-4, hidden-string-symm-1}; for recent work in this direction, see also\cite{yoshida-yangian}. The gauge-gravity correspondence tells us that the same symmetries must be present at the level of the conformal field theory, and indeed,  Yangian symmetry was shown to be present in the weak coupling limit of $SYM$ in \cite{witten-yangian-1, witten-yangian-2}. But  at the present time, it in not clear  how to extend this to the full gauge theory. One can however take the Dilatation operator to be the effective Hamiltonian for the gauge theory. Indeed in radial quantization of the conformal field theory, it $\it{is}$ the Hamiltonian. Moreover, the occurrence of Yangian symmetries has been a persistent theme in the literature on quantum spin chains\cite{pasq, ber-fel, ge-rev}. Hence it is natural to study the relation between the Yangian symmetries of the dilatation generator and those of the effective string actions. This analysis is one of the main points of the present paper. 

In doing this analysis, we emphasize the fact that the Dilatation operator of $\cal N$$=4$ SYM is a Hamiltonian matrix model. This is not surprising as the Dilatation generator is nothing but a particular  dimensional reduction of the full gauge theory on $R\times S^3$ \cite{bmn, plefka-matrix-1, plefka-matrix-3, plefka-matrix-2, janik-1, okuyama-1}. Indeed, the dimensional reduction of any non-Abelian gauge theory to one dimension will generate a quantum mechanical matrix model. In the context of superconformal Yang-Mills theory, this approach has been stressed in several papers. For example, in \cite{bmn} the relation of the Dilatation operator to a system of coupled Cuntz oscillators\cite{Gross-Gopakumar} was stressed. In \cite{jevicki-sft-1, jevicki-sft-2} progress has been achieved in deriving the pp wave string field theory Hamiltonian  by using a collective field theory approach towards the study of multi-matrix models. More recently, \cite{david-b} have also elaborated on this connection. Also, from the point of view of the connection to the pp wave limit of matrix theory \cite{plefka-matrix-1, plefka-matrix-3, plefka-matrix-2, keshav-et-al-1} the matrix model interpretation of the radial Hamiltonian of the gauge theory is quite natural.  For related recent work, see also \cite{volovich-1, volvich-2, bellucci-2}. The connection to quantum spin chains has to do with taking the Large N limit of these matrix model, while preserving normal ordering \cite{rajeev-lee-prl, rajeev-lee-normal}. Typically, while analyzing the large N limit of matrix models which are invariant under some global 'gauge group' ((U(N) in our case), one attempts to change variables to the gauge invariant observables. This can be achieved very explicitly, when only a single matrix is present. There are at least two well known methods of doing this. The first one has to do with a direct change of variables from the matrix elements to the eigenvalues\cite{Brezin}. The effective theory for the eigenvalues can then be mapped to one of free Fermions. The other method, which is related to the first one, has to do with the Bosonization of this theory of fermions through the approach of collective field theory\cite{Collective1, Collective2, Collective4}. By now there is a large literature devoted to these techniques and their applications to low dimensional string theory. When dealing with the quantum mechanics of several matrices, it is not clear how to make such direct approaches work (although extending the collective field theory formalism  has been shown to hold a lot of  promise\cite{Abhi2,jevicki-sft-1, jevicki-sft-2}). However, one can proceed to isolate the operators and the states that dominate the large N limit, work out the algebra of observables obeyed by the dominant observables and continue to work within this restricted sector. Doing this leads to a map between matrix models and spin chains. One sees that the observables that do dominate in the large N limit, which are basically nothing but normal ordered single trace operators, and the corresponding single trace states  can be mapped to observables  and states of quantum spin chains. In a sense the quantum spin chains play the same role in the case of Hamiltonian multi matrix models that the free Fermions do for the single matrix case. Hence, one may expect that the algebra of observables of the large N limit of multi matrix models is, in a sense, a generalization of the $W_{\infty}$ algebra. This is indeed true, and this Lie algebra , which was worked out in \cite{rajeev-lee-prl, rajeev-lee-normal, rajeev-lee-review}, will be the basic tool that we shall use to analyze the Yangian symmetries of the dilatation operator.  

One of the advantages to working within the matrix model framework in the study of Yangian symmetries is the following. It is well known that  for spin chains of finite length, the Yangian charges are not truly conserved. When one computes the commutator of the Yangian generators with the Hamiltonian, the commutators fail to vanish, and this failure has to do with effects that arise due to the finite size of the spin chains. We shall be able to show that such boundary terms have a very natural meaning in the Lie algebra of matrix model observables. They can be described as elements of a proper ideal of the  Lie algebra. This understanding allows us to look for modified definitions of the Yangian generators which are truly conserved. Such generators are presented in the section on the $SU(n)$ Yangian. They are nothing but the matrix elements  of the transfer matrix. This  allows one to realize the Yangian as a true symmetry of the matrix model/spin chains, even when the interactions are extremely short ranged, and only (though not necessarily) short states are being considered. Also, from a computational point of view, the matrix model computations seem to organize themselves in a way that makes the analysis of various conservation laws quite transparent.  

The other advantage of this formalism lies in motivating the continuum limit as a classical theory. By recasting the computation of the spectrum as a variational problem for the matrix model, we see that the passage to the continuum sigma model, can be motivated without having to invoke a long wavelength expansion. Also, this particular way of taking the continuum limit avoids the explicit reference to the spin coherent states, enabling us to carry out the analysis in general for $SU(n)$ and $SO(n)$, without having to construct spin coherent states for each value of $n$.  The contraction of the Hopf algebraic  Yangian symmetry to the semiclassical Yangian (which is a Lie Poisson symmetry) is also quite transparent in this language.

The organization of the paper is as follows. In the first section, the basic matrix model formalism is reviewed. In this section we pay special attention to the Lie algebra of normal ordered matrix model observables, and elaborate on the comments made above, about the relation of the boundary terms to a proper ideal of this algebra. Following this we analyze the one loop dilatation operator corresponding to the $SU(n)$ sectors, from the point of view of Yangian symmetries. We present the modified Yangian generators which are conserved, irrespective of the length of the states being considered. This is followed by an analysis  the Yangian invariance at higher loops, where we present the Yangian generators that commute with the two loop $SU(2)$ dilatation operator. We then go on to study the   Yangian symmetries of the full $SO(n=6)$ invariant one loop dilatation operator for the scalars. In this case, we work out the relation of the Yangian generators to the expansion of the transfer matrix. We also study the conservation properties of the Yangian generators. Following this, we analyze the continuum limit of these matrix models. The $SU(n)$ case leads to the well studied cases of  Heisenberg like integrable models, and we comment on the contraction of the Yangian symmetry of the matrix models to the semi-classical Yangian symmetry present in the continuum sigma models. Extending the   work reported in \cite{tseytlin-2, tseytlin-1}, we also present the sigma model arising in the $SO(n)$ case. Our sigma model differs in some details from the one presented in \cite{tseytlin-2, tseytlin-1}.In this case too, give a zero curvature representation for its equations of motion.  We then proceed to use the monodromy matrix of the sigma model to work out its semi-classical  $SO(n=6)$ Yangian symmetry, and discuss its implications on connecting the integrable structures present in the $SO(n)$ sector of SYM and its semi-classical string dual. 

\section{Matrix Model Techniques and Details:}
In this section we shall briefly review the basic matrix model techniques required for the analysis of the Dilatation generator. 

As is well known by now \cite{beisert-et-al-conformal, bmn}, the connection between matrix models and the dilatation generator of $\cal N$$=4$ SYM arises in the following way. The Dilatation generator can be interpreted as  a dimensional reduction of the gauge theory on $R\times S^3$. More precisely, it can be thought of as the effective quantum mechanical Hamiltonian obtained upon integrating out the higher Kaluza-Klien modes coming from the expansion of the degrees of freedom of the gauge theory in the spherical harmonics on $S^3$. The Conformal field theory being a theory of non-Abelian  
fields in the adjoint representation of the gauge group, which we shall take to be $U(N)$, produces a 
quantum mechanical theory of interacting matrices as its reduction to one dimension\cite{bmn, beisert-et-al-bmn, beisert-et-al-conformal}. At one loop the Dilatation generator is 'block diagonal', so it makes sense to  isolate the scalars and study their mixing among each other. The Fermions and the field strengths continue to run in loops and produce effective interactions, which were not present in the original gauge theory Hamiltonian. Or in other words, the dimensional reduction is not the naive one. The one loop dilatation operator in the scalar sector is the following matrix model \cite{ beisert-et-al-conformal}.
\beq
\Gamma_{SO(6)} = \frac{\lambda }{32\pi ^2N}Tr\left(2\left[a^{\dagger i},a^{\dagger j}\right]\left[a_j,a_i\right] - \left[a^{\dagger i},a_j\right]\left[a^{\dagger i},a_j\right]\right) \label{dilso6}.
\eeq
The flavor indices $i,j$ go from 1 to 6, and the Hamiltonian has a manifest $SO(6)$ (the R symmetry of the gauge theory) invariance. 
 The matrix creation and annihilation operators satisfy the standard commutation relations,
\beq
\left[ a^{ \alpha}_{i\beta }, a^{\dagger k\gamma }_{\delta} \right] = \hbar \delta_i^{k}\delta ^{\gamma}_{\beta} \delta ^{\alpha }_{\delta }.
\eeq 
The greek letters denote  $U(n)$ color indices\footnote{We have chosen to display $\hbar $ explicitly in the commutation relations to remind ourselves of the fact that it is a natural deformation parameter in the matrix model, in addition to $\frac{1}{N}$\cite{abhi-N=4-1}. Also in performing actual calculations $\hbar $ serves as a good book keeping device.}. 

\begin{figure}
\centerline{\epsfxsize=4.truecm \epsfbox{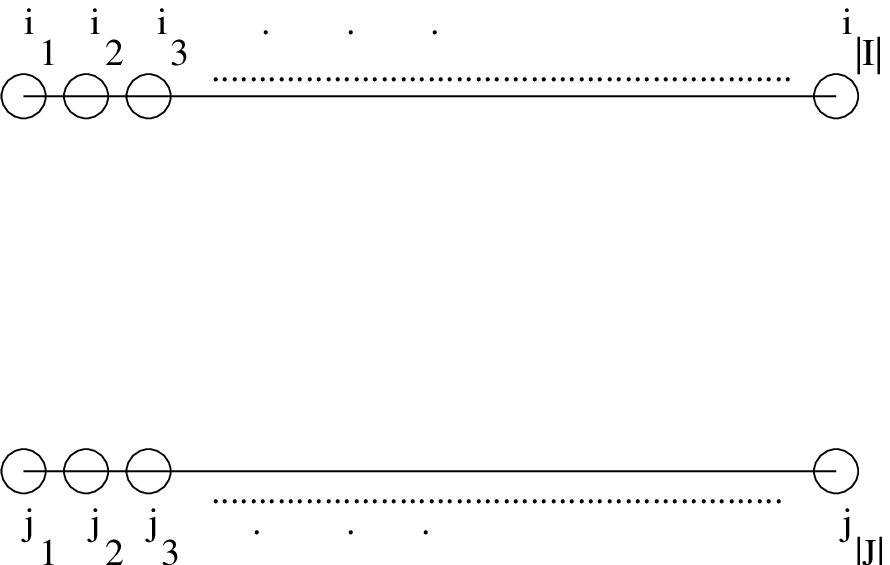}}
\caption{$\Theta ^{i_1 \cdots i_{|I}}_{j_1 \cdots j_{|J|}}$}
\label{tensor}
\end{figure}

\begin{figure}
\centerline{\epsfxsize=2.truecm \epsfbox{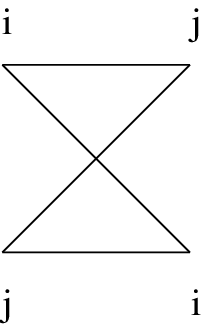}}
\caption{$\Theta ^{ij}_{ji}$}
\label{one-loop}
\end{figure}

\begin{figure}
\centerline{\epsfxsize=2.truecm \epsfbox{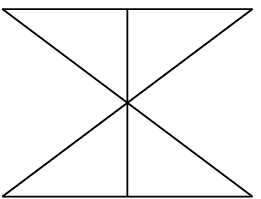}}
\caption{$\Theta ^{ikj}_{jki}$}
\label{two-loops}
\end{figure}

At higher loops it does not make sense to study the full bosonic sector by itself, as it is not closed. 
There does however exist a closed $SU(2)$ which is closed to all orders in perturbation theory. 
In the $SU(2)$ sector of the scalars,  the second term of the above Hamiltonian is absent, while the flavor indices take on two values. At the one loop level the corresponding matrix model Hamiltonian is,
\beq
\Gamma_{SU(2)} = \frac{\lambda }{16\pi ^2N}Tr\left(\left[a^{\dagger i},a^{\dagger j}\right]\left[a_j,a_i\right]\right) \label{dilsu2}.
\eeq
As has been out lined in several places in the literature, see for example \cite{beisert-et-al-conformal},  interpreting the dilatation generator as the Hamiltonian of a dynamical system allows one to understand operator mixing of the gauge theory as follows. Gauge theory operators correspond to states of the matrix model, for example, a generic multi trace operator built out of the adjoint scalars   
\beq
\Upsilon ^{I,J,\cdots, M} = Tr(\Phi ^{i_1} \cdots \Phi ^{i_{|I|}})Tr(\Phi ^{j_1} \cdots \Phi ^{j_{|J|}})\cdots Tr(\Phi ^{m_1} \cdots \Phi ^{m_{|M|}}),
\eeq 
maps to the state,
\beq
\Upsilon ^{I,J,\cdots, M} \mapsto \ket{I,J, \cdots, M} = O^I O^J \cdots O^M\ket{0},
\eeq
 where,
\beq
O^I = \frac{1}{\sqrt{N^{|I| -2}}}Tr \left(a^{\dagger i_1 }\cdots a^{\dagger i_n }\right).
\eeq
We have denoted ordered strings of indices by capital letters, for example, \\$\{i_1, i_2, \cdots ,i_{|I|}\}$ $= I$, while $|I|$ denotes the number of bits present in the string.
The action of the matrix model Hamiltonian on such states will in general produce a linear combination of multi trace states, which correspond to the operators that the original one mixes with. Or more precisely, the Callan-Symanzyk equation for the gauge theory can be recast as the Hamiltonian evolution equation for the matrix model,
\beq
i\hbar \frac{\partial }{\partial t}\ket{I_i \cdots I_n} = \Gamma \ket{I_i \cdots I_n}.,
\eeq
with 'time' playing the role of the logarithmic scale of the gauge theory. 
This basic formalism works for any $N$. However, we are interested in the large $N$ limit of the dilatation operator, which, from the above discussion amounts to studying the large $N$ limit of Hamiltonian multi-matrix models. It is not clear if methods of the kind that are employed in the study of single matrix models (such as carrying out a direct change of variables to the eigenvalues) have any obvious generalization to the multi-matrix cases. In the absence of such methods, one can nevertheless make progress. As was mentioned in the introduction, one starts out by isolating a complete set of states needed to describe this limit.  As one might expect, these are the single trace states. The overlap of single and multi-trace states is lower order in $\frac{1}{N}$. So the Hilbert space of the large $N$ theory consists of cyclically symmetric states of the kind,
\beq
|i_1 \cdots i_n> = |I> = \frac{1}{\sqrt{N^{|I| -2}}}Tr \left(a^{\dagger i_1 }\cdots a^{\dagger i_n }\right)|0>
\eeq
The dominant observables are the ones which (to leading order in  $\frac{1}{N}$) do not split single trace states into multi-trace ones. A little though shows that these observables are the ones for which  
normal ordering (in the sense of operator ordering) is compatible with the ordering implied by matrix multiplication. These are operators of the form,
\beq
\Theta ^{i_1 \cdots i_{|I}}_{j_1 \cdots j_{|J|}} = \Theta ^I_J = \frac{1}{\sqrt{N^{|I| +|J| -2}}}Tr \left(a^{\dagger i_1 }\cdots a^{\dagger i_{|I|} }a_{j_{|J|} }\cdots a_{j_1 }\right).
\eeq
Note the reverse order of the lower string in the definition of $\Theta ^I_J$.  It is useful to represent these tensors diagrammatically. One can denote the upper and lower set of indices by two horizontal lines e.g. fig-(\ref{tensor}). Contraction of the indices can be represented by lines connecting the two horizontal lines. For example, the tensor $\Theta ^{ij}_{ji}$, which is basically the one loop dilatation operator in the $SU(2)$ sector can be represented as, fig-(\ref{one-loop}). Clearly, for $SU(n)$ scalars, it is not necessary to depict the indices in the diagrams, as the way the horizontal and vertical lines are contracted carries all the information necessary for specifying the tensor. For example, $\Theta ^{ijk}_{kji}$, which is part of the two loop dilatation operator, in the same $SU(2)$ sector, can be specified by fig-(\ref{two-loops}). From now on , we shall only depict the un contracted indices explicitly in these diagrams. 

The matrix model operators act on the large $N$ states as follows,
\beq
\Theta ^I_J |K> = \delta ^K_{JA}|IA>,
\eeq 
or to put it in words, the operator $\Theta ^I_J $, checks if the string making up the state $|K> $ can be split in two parts in such a way that the first part is equal to $J$. If it is possible, then that part of the string is replaced by $I$. 
By looking at the action of these operators on single trace states, one sees that they define a Lie algebra. We shall refer to this as the planar Lie algebra $Pl(n)$, (here $n$ stands for the number of matrices present in the problem), and it was worked out in \cite{rajeev-lee-prl}.  
\beq
\left[\Theta ^I_J, \Theta ^K_L\right] = g^{IKM}_{JLN}\Theta ^N_M.
\eeq

The structure constants can be read off from the explicit form of the Lie bracket given below.
\beqs
\poisn{\Theta ^I_J}{\Theta ^K_L }=(\hbar )^{|K|}(\delta ^K_ J\Theta ^I_L + \sum_{J_1J_2 =J}\delta ^K_ {J_2}\Theta ^I_{J_1L} + \sum_{J_1J_2 =J}\delta ^K_ {J_1}\theta ^I_{LJ_2}\nonumber \\
+\sum_{J_1J_2 J_3=J}\delta ^K_ {J_2}\Theta ^I_{J_1LJ_3} + \sum_{\stackrel{J_1J_2 J_3=J}{K_1K_2=K}}\delta ^{K_1}_ {J_3}\delta ^{K_2}_ {J_1}\tilde{\Theta }^I_{J_2L})\nonumber \\
+(\hbar )^{|J|}(\sum_{K_1K_2 =K}\delta ^{K_1}_{J}\Theta ^{IK_2}_{L} + \sum_{K_1K_2 =K}\delta ^{K_2}_{J}\Theta ^{K_1 I}_{L} + \sum_{K_1K_2K_3 =K}\delta ^{K_2}_{J}\Theta ^{K_1IK_3}_{L}\nonumber\\
+\sum_{\stackrel{J_1J_2=J}{K_1K_2=K}}\delta ^{K_1}_ {J_2}\delta ^{K_2}_ {J_1}\tilde{\Theta }^I_{L} + \sum_{\stackrel{J_1J_2=J}{K_1K_2K_3=K}}\delta ^{K_1}_ {J_2}\delta ^{K_3}_ {J_1}\tilde{\Theta }^{IK_2}_{L})\nonumber \\
+\sum_{\stackrel{J_1J_2=J}{K_1K_2=K}}(\hbar )^{|K_1|}\delta ^{K_1}_ {J_2}\Theta ^{IK_2}_{J_1L} + \sum_{\stackrel{J_1J_2=J}{K_1K_2=K}}(\hbar )^{|K_2|}\delta ^{K_2}_ {J_1}\Theta ^{K_1I}_{LJ_2}\nonumber \\
+\sum_{\stackrel{J_1J_2J_3=J}{K_1K_2K_3=K}}\hbar ^{J_1 + J_3}\delta ^{K_1}_ {J_3}\delta ^{K_3}_ {J_1}\tilde{\Theta }^{IK_2}_{J_2L} - \chose{I \Leftrightarrow K}{J \Leftrightarrow K}. \label{planar-lie}
\eeqs

In the above equation, 

\beq
\tilde{\Theta }^{I}_{J} = \Theta ^I_J - \sum_{i=1}^{n} \Theta ^{kI}_{kJ}.
\eeq
Although the Lie algebra looks rather complicated when written out in this fashion, all the terms above have rather simple diagrammatic interpretations, for which we shall refer the reader to \cite{rajeev-lee-review}. 
\subsection{Boundary Terms:} If one discards, the terms involving $\tilde \Theta $ from the above equation, one can see that the terms involving only the $\Theta $ s still continue to form a  Lie algebra. In-fact the 
$\tilde \Theta $'s span a proper ideal $K(n)$ of the planar Lie algebra {\footnote {A more detailed account of the ideal generated by these terms is given in the appendix}}. These terms have the interesting property of being able to act as  Weyl operators on the single trace states,
\beq
\tilde \Theta ^I_J |K> = \delta ^K_J|I>,
\eeq
and they obey the following commutation relation, 
\beq
\left[\tilde \Theta ^I_J,  \tilde \Theta ^K_L\right] = \delta ^K_J\tilde \Theta ^I_L - \delta ^I_L\tilde \Theta ^K_J.
\eeq 
It is important to note that these operators have more than one representation in terms  of the $\Theta $s. For example,

\beq
\tilde{\Theta }^{I}_{J} = \Theta ^I_J - \sum_{i=1}^{n} \Theta ^{kI}_{kJ} = \Theta ^I_J - \sum_{i=1}^{n} \Theta ^{Ik}_{Jk}.
\eeq
This implies that the $\Theta $'s are not linearly independent. The most obvious relation implied among them, from the equation above, is that
\beq
\Theta^{IK}_{JK} = \Theta ^{KI}_{KJ} 
\eeq
\begin{figure}
\centerline{\epsfxsize=4.truecm \epsfbox{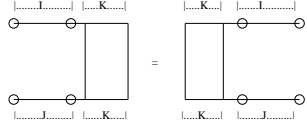}}
\caption{$\Theta^{IK}_{JK} = \Theta ^{KI}_{KJ}$}
\label{relation}
\end{figure}
where, by {KI}, we mean the concatenation of the strings $K$ and $I$, $KI = \{i_1\cdots i_{|I|}k_1 \cdots k_{|K|}\}$ fig-(\ref{relation}). This relation between the matrix model observables is going to prove quite useful in   doing the calculations that we shall present later in the paper . 
The origin of these terms lies in the fact the in looking at the action of the matrix model observables   on states of  finite lengths, the cyclicity of the trace produces terms which one might not have guessed naively to be there. \\ 
{\bf An example of the generation of a boundary term:} Let us look at the action of a product of $\chose{3}{3}$ and $\chose{2}{2}$ operators on a state of length three.
\beq
\Theta ^{i_1 i_2 i_3}_{j_1 j_2 j_3}\Theta ^{k_1 k_2}_{l_1 l_2}|a_1 a_2 a_3>
\eeq
The cyclicity of the trace is going to produce an action of the first operator on the bits  $a_3$ and $a_1$, resulting in,
\beq
\delta ^{a_3}_{l_1}\delta ^{a_1}_{l_2}\Theta ^{i_1 i_2 i_3}_{j_1 j_2 j_3}|k_2,a_2,k_1> = \delta ^{k_2}_{j_1} \delta ^{k_1}_{j_3}\left( \delta ^{a_2}_{j_2} \delta ^{a_3}_{l_1} \delta ^{a_1}_{l_2}\right)|i_1,i_2,i_3>,
\eeq
which can be written as,
\beq
\delta ^{k_2}_{j_1} \delta ^{k_1}_{j_3}\tilde{\Theta }^{i_1i_2i_3}_{j_2l_1l_2}\left(|a_2,a_3,a_1> \sim |a_1,a_2,a_3>\right). 
\eeq
Clearly, this term was produced be cause of two reasons, the finite size of the state  matching that of $J$, and the cyclicity of the trace. These boundary terms have to be treated separately. Generalizing to tensors of arbitrary length, we see then that in evaluating the product of two operators, $\Theta ^I_J \Theta ^K_L $, we shall have a term, $\sum_{\stackrel{J=J_1J_2J_3}{K=K_1K_2}}\delta^{K_1}_{J_3}\delta^{K_2}_{J_1}\tilde{\Theta }^{I}_{J_2L}$, which is the second term in the second line of (\ref{planar-lie}). Similar considerations lead to the other boundary terms in the Lie algebra as well. Keeping in mind the impending continuum limits, it is easy to convince oneself, that these boundary terms drop out, when one formally considers states of infinite length. Moreover, of the commutator of two matrix model operators can be written as a sum f the finite rank $\tilde \Theta $ operators, then one can safely set their commutator to zero in the continuum limit. As we shall see later, the commutator of the standard Yangian generators with the Hamiltonian are exactly of this sort, and we shall be able to find a new set of generators for which this problem does not exist even on states of finite size. 
\subsection{Spin Chains:}As outlined in previous papers \cite{rajeev-lee-prl, rajeev-lee-normal}, the connections to traditional quantum spin chains arises when one looks at operators $\Theta ^I_J$ for which $|I|=|J|$. These matrix model operators can be represented by spin operators $S^i_j(l)$, where $l$ is the lattice index. One can think of the states of the matrix model as those of the spin chains,
\beq
 |I> = \frac{1}{\sqrt{N^{|I| -2}}}Tr \left(a^{\dagger i_1 }\cdots a^{\dagger i_{|I|} }\right)|0> \sim |i_1, i_2 \cdots i_{|I|}>,
\eeq
where, the bits can be thought of as spins on a one dimensional lattice with periodic boundary conditions. The operators, $\Theta ^{i_1 \cdots i_n}_{j_i \cdots j_n}$ can be identified with sums of products of  spin operators,
\beq
\Theta ^{i_1 \cdots i_n}_{j_i \cdots j_n} \rightarrow \sum_lS^{i_1}_{j_1}(l)S^{i_2}_{j_2}(l+1)\cdots S^{i_n}_{j_n}(l+n-1), 
\eeq
where the spin operators commute at unequal sites and satisfy the Weyl relations,
\beq
S^i_j(k)S^p_q(k) = \delta ^p_jS^i_q(k)
\eeq
at the same lattice points. 

It is worth emphasizing a few points at this juncture. If one works out the commutation relations of the $\Theta$s using the representation in terms of the spin operators, then one would not get the boundary $\tilde \Theta $ terms. That is because these finite rank operators have no 'microscopic description', i.e in terms of spin operators at individual lattice sites. However, they will arise  when one computes the antisymmetric part of the product of the action such translation invariant spin chain operators on any state of a given length.  The other point has to do with the fact that when $|I| \neq |J|$ in $\Theta ^I_J$, the planar Lie algebra continues to make sense, however there is no realization of such matrix model operators in terms of spin matrices.  Dynamical systems having such terms in their Hamiltonians are of a lot of interest in the context of the study of  $\cal N$$=4$ SYM. For example,  the higher loop dilatation generators in the closed $su(2|3)$ sector on the sector exhibit this property \cite{beisert-dyn}. One might hope that the matrix model point of view might shed some new light on these 'dynamical' spin chains. 
    
 
\section{Non-Local Conserved Charges and Yangian Invariance of the Dilatation Generator: $SU(n)$ Sectors: }
In this section we shall work out the non-local conserved Yangian charges for the one loop $SU(2)$ part of the Dilatation operator. Ignoring the trivial piece, proportional to the identity, and setting $\frac{\lambda}{16\pi ^2}=1$, the, the Heisenberg Hamiltonian is,
\beq
D_2 = \sum_lP_{l,l+1} = \sum_lS^i_j(l)S^j_i(l+1) = \Theta ^{ij}_{ji}
\eeq
The Lax operator for this Hamiltonian, is well known \cite{fadbook,fad-rev}, and it takes on the form,
\beq
L_l = \symmat{u+iS^3(l)}{iS^-(l)}{iS^+(l)}{u-iS^3(l)}.
\eeq
As is evident, it is a matrix, each element of which is an operator on the $l$th site of the spin chain. $S^3$ is the third Pauli matrix, and $S^{\pm}$ are the spin rising and lowering operators, while  $u$ is the spectral parameter. It is useful to shift the spectral parameter by $\frac{i}{2}$, so that, the matrix elements of the Lax matrix take on the following simple form,
\beq
\left(L_n\right(u))_{ij} = \left(uI(n)\delta _{ij} + iS_{ji}(n)\right).
\eeq
The transfer matrix, is defined in terms of the Lax matrix in the usual way,
\beq
T(u)_{ba} = L_1^T(u)_{aa_1}L_2^T(u)_{a_1a_2}\cdots L_{J}^T(u)_{a_{J-1}b},
\eeq
where $J $ is the length of the chain, and $T$ denotes the transpose over the auxiliary space. 
By dividing out the Lax by the spectral parameter, one can get the transfer matrix in a form that lends itself to an expansion about $u =\infty$. 
\beq
T = \symmat{T_{11}(u)}{T_{12}(u)}{T_{21}(u)}{T_{22}(u)} = \sum_{n=0}^{\infty}u^{-n}T^n,
\eeq
 Each element of the matrix is an operator acting on the entire spin chain, and hence has a matrix model realization. It is more useful to think not in terms of the transfer matrix but its transpose $\tilde T$ which can be written out as.
\beq
\tilde T^a_b = (I + \frac{1}{u}S_1)^a_{a_1}(I + \frac{1}{u}S_2)^{a_1}_{a_2}\cdots(I + \frac{1}{u}S_n)^{a_{n-1}}_{b},
\eeq 
where we have used the upper and lower indices with the convention $(S_{k})^{i}_j = (S_{k})_{ij}$.  This matrix is completely equivalent to the transfer matrix,  so we shall drop the tilde's from now on.  Clearly, term by term in powers of $u$, one can translate the transfer matrix into matrix model operators. For example, the first two terms, which we shall interpret as the generators of the Yangian of $SU(n)$ are,
\beq
( T^1)^i_j = \Theta ^i_j
\eeq 
\beq
(T^2)^i_j = \sum_L \Theta ^{i L i_1}_{i_1 L j}.
\eeq
The matrix model equivalent of the full transfer  matrix is,
\beq
 T^q_b = I^a_b + \sum_{n=0}^\infty \frac{1}{u^n}\Theta ^{a {I_1}{i_1} {I_2}{i_2} \cdots {I_n}{i_n}}_{{i_1}{I_1}{i_2}{I_2}\cdots {i_n}{I_n}b},
\eeq
where a sum over the repeated strings is assumed. A few words about the validity of such infinite sums are in order. Although the operators representing the expansion of the transfer matrix are infinite sums of a  the kind $\sum_{I}\Theta ^{AIB}_{CID}$, when one looks at their action on any state $|K>$, of finite length,  only a finite number of such terms contribute. Namely, only the terms for which $|CID|\leq |K|$ will have non-zero action on such states. Which means that $||\sum_{I}\Theta^{AIB}_{CID} |K> ||^2 < \infty$, which allows us to consider such infinite sums. 

Coming back to the main discussion,  we can now
proceed to show that $T^a_b$ are exactly  conserved, without any problems arising due to the boundary terms. Moreover, the algebra generated by the $T^a_b$'s is equivalent to the $SU(N)$ Yangian. The second part of the statement is very easy to substantiate. The Yang-Baxter relations imply that
\beq
[(T^{n+1})^b_c, (T^{m})^a_d] - [(T^{n})^b_c, (T^{m+1})^a_d] + (T^{n})^a_c(T^{m})^a_d - (T^{m})^a_c (T^{n})^b_d =0 \label{ybe-comp}
\eeq
One can see that from this that $T^0$ lies in the center of the algebra, and that $T^n, n\geq3 $ can be found from the knowledge of the basic charges $T^1$ and $T^2$. and it is well known, that the associative algebra defined by (\ref{ybe-comp}), is finitely generated, i.e the first two generators are the only independent ones, and that it is equivalent to the Yangian of $SU(n)$  upto a redefinition of the generators\cite{ge-rev}. We shall also comment on how to carry out this redefinition to put the Yangian relations in a more conventional form. For algebraic details on how to do generate the full transfer matrix from the first two generators, we  shall refer to \cite{ge-rev}. From the discussion above,  we see that it is enough to prove that $T^1$ and $T^2$ are conserved. 

\begin{figure}
\centerline{\epsfxsize=8.truecm \epsfbox{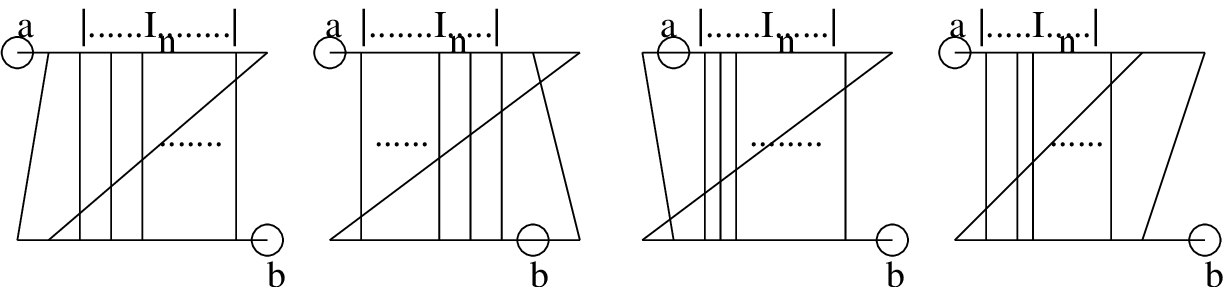}}
\caption{$F(n)^a_b$}
\label{Fn}
\end{figure}

$(T^1)^a_b$ is the generator of $SU(n)$ transformations, and it is easy to see that it is conserved. 
When one evaluates the commutator of $(T^2)^a_b$ with $H$, one will have three kind of terms. The first kind  are of order one in the deformation parameter $\hbar $, which we shall set equal to one at the end of the calculation. The second kind of terms are of order $\hbar ^2$, and finally we shall have the $\tilde{\Theta }$ terms. To start with, we write $(T^2)^a_b$ as,
\beq
(T^2)^a_b = \sum_{n=0}^\infty \Theta ^{aI_ni}_{iI_n b},
\eeq
where the subscript in $I_n$ denotes that $|I| = n$.
Now we can see that,
\beqs
\poisn{\Theta ^{ij}_{ji}}{\sum_{n=0}^\infty \Theta ^{aI_ni}_{iI_n b}} = \sum_n \hbar \left(\Theta ^{ai_1I_ni_2}_{i_1i_2I_nb} + \Theta ^{aI_ni_1i_2}_{i_2I_nbi_1} - \Theta ^{i_1aI_ni_2}_{i_2i_1I_nb} - \Theta ^{aI_ni_1i_2}_{i_1I_ni_2b}\right)\nonumber \\
+\hbar ^2\left( \Theta ^{i_1aI_{n-1}i_2}_{i_2i_1I_{n-1}b} + \Theta ^{aI_{n-1}i_1i_2}_{i_1I_{n-1}i_2b} - \Theta ^{ai_1I_{n-1}i_2}_{i_1i_2I_{n-1}b}- \Theta ^{aI_{n-1}i_1i_2}_{i_2I_{n-1}bi_1}\right),
\eeqs
while, the $\tilde{\Theta }$ terms vanish identically. So the commutator has the functional form,
\beq
\poisn{\Theta ^{ij}_{ji}}{\Theta ^{aI_ni}_{iI_n b}} = \hbar F^a_b(n) - \hbar^2 F^a_b(n-1),
\eeq
where, the function $F(n)^a_b$ has the diagrammatic representation as fig-(\ref{Fn}).
Also it is easy to see that, the $[\Theta ^{ij}_{ji}, \Theta ^{al}_{lb}]$ is of $O(\hbar ^2)$, i.e.  
\beq
F^a_b(-1)=0
\eeq
After setting $\hbar = 1$, and summing over $n$, one sees that
\beq
\poisn{H}{(T^2)^a_b} = 0.\label{comm-one-loop}
\eeq 
Because the $\tilde \Theta $ terms vanish, these charges continue to be true generators of symmetries for the matrix model. The usual problems with the Yangian charges failing to commute by boundary terms are not present in this way of thinking about the Hopf algebraic symmetry.
Now we even have a whole tower of exactly conserved charges. (\ref{ybe-comp})implies that iterated commutators of the first two charges generate the full transfer matrix, hence all the matrix elements of the transfer matrix are exactly conserved, i.e
\beq
\poisn{H}{(T^{n+1})^a_b}_{n\geq 0} = \poisn{H}{\Theta^{aI_1i_1I_2i_2\cdots I_ni_n}_{i_1I_1i_2I_2\cdots i_nI_nb}} =0
\eeq
Hence, we see that there are an infinite number of non-local conserved charges for the matrix model. The charges correspond to the terms in the un-traced transfer matrix. In summary, the modified generators for the Yangian of $SU(n)$ are,
\beqs
(Q^1)^a_b = \Theta ^a_b \nonumber \\
(Q^2)^a_b = \sum_I \Theta^{aIi}_{iIb}  
\eeqs
We do not quite have to verify the Serre relations (as we do for various other generators in what is to follow) because of the following reason. The Serre relations are nothing but the condition for the co-product of the Yangian algebra to define an algebra homomorphism, and it is known that the transfer matrix has such a co-product. So the interpretation of the Yangian generators as the matrix elements of the transfer matrix ensures the Serre relations. The co-product for the transfer matrix is,
\beq
\Delta T^a_b(u) = T^a_c(u) \otimes T^c_b(u),
\eeq  
which generates the co-product for our Yangian generators,
\beqs
\Delta (Q^1)^a_b(u) = (Q^1)^a_b(u) \otimes I + I\otimes (Q^1)^a_b(u)\nonumber \\
\Delta (Q^2)^a_b(u) = (Q^2)^a_b(u) \otimes I + I\otimes (Q^2)^a_b(u) + (Q^1)^a_c(u) \otimes(Q^1)^c_b(u).
\eeqs
\subsection{Conventional Presentation of the Yangian Algebra and Conservation Laws:}
To make contact with the more conventional forms of the Yangian generators, we consider the algebra   generated by,
\beqs
(Q^1)^a_b = \Theta ^a_b, \nonumber \\
(Q^2)^a_b = \sum _{I_i \neq \Phi}\Theta ^{aI_1i_1}_{i_1I_1b} - \Theta ^{i_1I_1a}_{bI_1i_1}
\eeqs
The first charge is the conventional $SU(n)$ generator, while as a spin chain operator, the next charge is equivalent to,
\beq
(Q^2)^a_b = \sum_{i<j}\left(S^a_k(i)S^k_b(j) - S^k_b(i)S^a_k(j)\right)
\eeq
These are the forms of the generators that are conventionally employed. While, these have the same amount of information as the previous generators, these are not conserved on chains of finite size. 
It is easy to see that,
\beqs
\poisn{(Q^1)^a_b}{(Q^1)^c_d} = \delta ^c_b(Q^1)^a_d - \delta ^a_d(Q^1)^c_b \nonumber \\
\poisn{(Q^1)^a_b}{(Q^2)^c_d} = \delta ^c_b(Q^2)^a_d - \delta ^a_d(Q^2)^c_b.\label{yan-psn1}
\eeqs 

The next relation is more interesting, 
\beqs
\poisn{(Q^2)^a_b}{(Q^2)^c_d} = \delta ^c_b(Q^3)^a_d - \delta ^a_d(Q^3)^c_b + \nonumber \\
+ \sum _{I_1,I_2 \neq \Phi} \Theta ^{aI_1cI_2i_1}_{dI_1i_1I_2b} + \Theta ^{cI_1aI_2i_1}_{i_1I_1dI_2b} + \Theta ^{cI_1i_1I_2a}_{i_1I_1bI_2d}\nonumber \\
+\Theta ^{aI_1i_1I_2c}_{dI_1bI_2i_1} + \Theta ^{i_1I_1aI_2c}_{bI_1dI_2i_1} + \Theta ^{i_1I_1cI_2a}_{bI_1i_1I_2d} - \chose{c \Leftrightarrow a}{d \Leftrightarrow b }.
\eeqs
$(Q^3)^a_d$ is new charge. 
\beq
(Q^3)^a_b = \Theta ^{aI_1i_1I_2i_2}_{i_1I_1i_2I_2b} - \Theta ^{i_1I_1i_2I_2a}_{bI_1i_iI_2i_2}
\eeq
More importantly, we note  (by say, looking at its action on single trace states) that,  the other terms are equivalent to, a cubic combination of the first charge. 

\beqs
\sum _{I_1,I_2 \neq \Phi} \Theta ^{aI_1cI_2i_1}_{dI_1i_1I_2b} + \Theta ^{cI_1aI_2i_1}_{i_1I_1dI_2b} + \Theta ^{cI_1i_1I_2a}_{i_1I_1bI_2d}\nonumber \\
+\Theta ^{aI_1i_1I_2c}_{dI_1bI_2i_1} + \Theta ^{i_1I_1aI_2c}_{bI_1dI_2i_1} + \Theta ^{i_1I_1cI_2a}_{bI_1i_1I_2d} - \nonumber \\ \chose{c \Leftrightarrow a}{d \Leftrightarrow b } \equiv (Q^1)^a_d (Q^1)^c_e(Q^1)^e_b - (Q^1)^a_e(Q^1)^e_d(Q^1)^c_b
\eeqs

So one can write the Poisson bracket between the $Q^2 $'s as 
\beq
 \poisn{(Q^2)^a_b}{(Q^2)^c_d} = \delta ^c_b(Q^3)^a_d - \delta ^a_d(Q^3)^c_b + (Q^1)^a_d (Q^1)^c_e(Q^1)^e_b - (Q^1)^a_e(Q^1)^e_d(Q^1)^c_b
\eeq

This implies that, one can eliminate $Q^3$ and obtain a Serre relation between the first two charges, which is,
\beqs
\poisn{(Q^1)^a_b}{\poisn{(Q^2)^c_d}{(Q^2)^e_f}} - \poisn{(Q^2)^a_b}{\poisn{(Q^1)^c_d}{(Q^2)^e_f}} = \nonumber \\ 
\poisn{(Q^1)^c_b}{(Q^1)^c_f(Q^1)^e_p(Q^1)^e_d -(Q^1)^c_p(Q^1)^p_f(Q^1)^e_d}\nonumber \\
-\delta ^e_d\left( (Q^1)^a_f(Q^1)^c_p(Q^1)^p_b - (Q^1)^a_p(Q^1)^p_f(Q^1)^c_b\right)\nonumber \\
+\delta ^c_f\left( (Q^1)^a_d(Q^1)^e_p(Q^1)^p_b - (Q^1)^a_p(Q^1)^p_d(Q^1)^e_b\right)\label{yan-psn2}
\eeqs
The algebra generated by (\ref{yan-psn1}, \ref{yan-psn2})is the $SU(n)$ Yangian.
The breakdown of conservation of these charges, due to the finite size of the states, has the simple characterization as the their commutator with the Hamiltonian belonging to the ideal generated by the $\tilde \Theta$ terms. This, it turns out, is the case with the conventional presentation of the Yangian generators. Let us investigate this in some detail. It is very easy to see that $(Q^1)^a_b$ is conserved. The first part of $(Q^2)^a_b$ is nothing but $T^2$ and it is conserved.
\beq
(Q^2)^a_b = (T^2)^a_b - \Theta ^{i_1I_1a}_{bI_1i_1},
\eeq
 After forming the antisymmetric combination $(Q^2)^a_b$ one sees that it is no longer conserved on states of finite size, i.e its conservation is violated by finite rank operators. 
 One can check, that in evaluating the commutator of the second term with $H$, all the $\Theta $ terms vanish and that we are left with the result,
\beq
\poisn{H}{(Q^2)^a_b} = -\sum_n \left(\tilde{\Theta }^{i_1aI_n}_{bI_ni_1} - \tilde{\Theta }^{i_1I_na}_{I_nbi_1}\right)
\eeq
Hence, the conservation of $(Q^2)$ is violated by finite rank operators, but, these terms drop out in the continuum limit. The simplest way to see that is as follows. On states of infinite length, one can formally write 
\beq
(Q^2)^a_b = (T^2)^a_b - \frac{1}{2} (Q^1)^a_d(Q^1)^d_b + (Q^1)^a_b.
\eeq

In this form, the conservation of $(Q^2)$ follows from that of $(T^2)$ and $(Q^1)$. However, 
this expression for $(Q^2)$ is not valid on states of any finite length, as boundary terms are going to violate it, which is the origin of the $\tilde{\Theta }$ terms in the algebra. As a simple illustration, one can see that, the second part of  $(Q^2)$, which is the origin of the violation of its conservation; i.e. $-\sum_n \Theta ^{iI_na}_{bI_ni}$ contains among other terms the term $-\Theta ^{ija}_{bji}$, corresponding to the case $n=1$. Now, if one looks at is action on a state of length four, $\ket{j_1 \cdots j_4}$, it will produce,
\beq
-\Theta ^{ija}_{bji}\ket{j_1 \cdots j_4} = -\delta ^{j_3}_b \ket{a j_2 j_1 j_4} + \mbox{ other terms }.
\eeq
Such terms are never generated by $(Q^2)$ in the form above. However, these terms are only present on states of a finite size, where the cyclicity of the trace induces effective interactions between certain sites (the first and the third in this case), which are not present when one formally looks at states of infinite length.  

{\bf{The special case of $SU(2)$:}} In the case of $SU(2)$, we can encode the same information given above, in the following form. We introduce,
\beq
(j^1)^i = \Theta ^a_b (\sigma ^i)^b_a
\eeq
and
\beq
(j^2)^i =\sum_{I} \epsilon ^{ijk}\left( \Theta ^{aIc}_{bId} (\sigma ^j)^b_a(\sigma ^k)^c_d )\right),  
\eeq
where the $\sigma $s are the Pauli matrices, with the convention that $(\sigma ^k)^c_d \equiv (\sigma ^k)_{dc}$.

One can now check by explicit computations that the following equations are satisfied. 
\beq
\poisn{(j^1)^i}{(j^1)^j} = \epsilon ^{ijk} (j^1)^k, \nonumber 
\eeq
\beq
\poisn{(j^1)^i}{(j^2)^j} = \epsilon ^{ijk} (j^2)^k, \nonumber 
\eeq
\beqs
\poisn{(j^2)^i}{\poisn {(j^2)^j}{(j^1)^k}} - \poisn{(j^1)^i}{\poisn {(j^2)^j}{(j^1)^k}} = C_{ijkabc}\{(j^1)^a, (j^1)^b, (j^1)^c \}\nonumber \\
\poisn{\poisn{(j^2)^i}{(j^2)^j}}{\poisn{(j^1)^k}{(j^2)^l}} + \poisn{\poisn{(j^2)^k}{(j^2)^l}}{\poisn{(j^1)^i}{(j^2)^j}} = \nonumber \\
\left(C^{ijrabc}\epsilon ^{klr} + C^{klrabc}\epsilon ^{ijr}\right) \{(j^1)^a, (j^1)^b, (j^2)^c \},
\eeqs
where,
\beqs
C_{abcpqr} = \frac{1}{24} \epsilon _{amn}\epsilon _{bqm}\epsilon _{clt}\epsilon _{nlt},\nonumber 
\eeqs
and $\{x,y,z\}$ denotes the totally symmetrized combination. This  definition of the Yangian algebra is the one used in the study of $SU(2)$ spin chains. Although it has the advantage of making the underlying $SU(2)$ structure quite explicit, from the discussion above, we note that the symmetries generated by these generators become exact only in the continuum limit.

\subsection{Yangian Invariance at Two Loops:} In this section we study, the commutation relation of the Yangian generator with the two loop Dilatation operator in the $SU(2)$ sector. As one might expect,  at two loops, the Yangian generators as they stand,  do not commute with the Dilatation operator, however,  their failure to commute can be captured in a rather simple way. This also allows us to find the deformed Yangian generators, which do commute with the two loop dilatation generator upto terms of $O(\lambda ^3)$. The analysis also useful in constructing an exact  non-local conserved charge for the two loop Hamiltonian 
\beq
D_4 = -6L + 8\sum_iP_{i,i+1} - 2 \sum_iP_{i,i+2} = -6\Theta ^i_i + 8\Theta ^{ij}_{ji} -2\Theta ^{ijk}_{kji}
\eeq
 As is evident from the formula for the two loop dilatation generator, it is a spin chain with next to nearest neighbor interactions. Although it is not clear how one might try to generalize an approach like the algebraic Bethe ansatz to the spin chains obtained from higher loop analysis of $SYM$,  symmetry considerations have been utilized in solving spin chains with long range interactions in the past. The classic example of this is the case of the Haldane-Shastry long ranged spin chain \cite{hal-shas}. Indeed for most quantum spin chains, finding the underlying Hopf algebraic symmetry and constructing the corresponding $R$ or transfer matrix for the problem are very nearly the same things. Although the higher loop dilatation operators are not understood at this level of mathematical completeness, for the two loop operator in particular, certain higher local charges have been generated based on the consideration of parity pairs \cite{beisert-et-al-conformal}.  Moreover, there is also evidence to support that this particular operator is one of the terms in the expansion of the Inozemtsev long ranged spin chain, for which, the Yangian is indeed a true symmetry \cite{long-range-inz}. Hence, though there is no particular reason to believe that the $SU(n)$ Yangian is an exact   symmetry at the two loop level, the fact that the Inozemtsev long range spin chain is capable of describing the two loop Hamiltonian, implies that it should be able to modify the Yangian charges (in a way that is consistent with the analysis presented in \cite{long-range-inz}) so that they continue to commute with the dilatation generator upto terms of $O(\lambda ^3)$. We shall see below that this is indeed the case. Moreover,  it has recently been shown that one can incorporate the two loop correction to the dilatation operator by modifying the Bethe equations that follow from the one loop Hamiltonian \cite{minahan-et-al-su2}. So the Yangian symmetry cannot be violated at the two loop level in some intractable fashion. Hence as a starting point,  it is important to not by how much the Yangian fails to be a symmetry at two loops. In what is to follow, we shall not bother about the $\tilde \Theta $ terms, and work within $Pl(n)/K(n)$. 
  
It is easy to see that,
\beq
\left[\Theta ^{pqr}_{rqp}, (Q^1)^a_b\right] =0;
\eeq
i.e the first generator of the $SU(n)$ Yangian commutes with the two loop Hamiltonian. The commutation of the second generator requires the evaluation of,
\beq
\left[\Theta ^{pqr}_{rqp}, (Q^2)^a_b\right] = \left[\Theta ^{pqr}_{rqp}, \sum_n\left(\Theta ^{aI_ni}_{iI_nb} - \Theta ^{iI_na}_{bI_n i}\right)\right]
\eeq
Our starting point (as in the one loop case) is the commutation relation of the two loop Hamiltonian with the operator,
\beq
\sum_n\Theta ^{aI_ni}_{iI_nb} \equiv \sum_{p<q}S^a_i(p)S^i_b(q)
\eeq
As in the previous case, $I_n$ denotes a sequence of $n$ 'bits'. $I_n = \{i_1 i_2 \cdots i_n\}$. In what is to follow we shall start out assuming $n\geq 2$, the cases with smaller values of $n$ are a little subtle and shall be treated separately. We shall not display the $\tilde {\Theta }$ terms explicitly. Those correspond to the boundary terms and shall drop out in the limit of large lengths. We shall only concentrate on the commutator of the term involving next to nearest neighbor interactions. By using the planar Lie algebra, one sees that,
\beqs
\left[\Theta ^{pqr}_{rqp}, \Theta ^{aI_ni}_{iI_nb}\right] = \hbar \left(\Theta ^{ajkI_ni}_{kjiI_nb} + \Theta ^{aI_njki}_{iI_nbkj} - \Theta ^{jkai_ni}_{ikjI_nb} - \Theta ^{aI_nijk}_{iI_nkjb} \right) \nonumber \\ \hbar ^3\left(\Theta ^{jkaI_{n-2}i}_{ikjI_{n-2}b} +\Theta ^{aI_{n-2}ijk}_{iI_{n-2}kjb}  - \Theta ^{ajkI_{n-2}i}_{kjiI_{n-2}b} - \Theta ^{aI_{n-2}jki}_{iI_{n-2}bkj}\right). 
\eeqs
All terms of $O(\hbar ^2)$ vanish. Or in other words, denoting the terms within the parentheses on the right hand side by $F^a_b(n)$, the equation above takes the following functional form.
\beq
 \left[\Theta ^{pqr}_{rqp}, \Theta ^{aI_ni}_{iI_nb}\right] = \hbar F^a_b(n) - \hbar ^3 F^a_b(n-2) 
\eeq
Similarly, we also have to look at,
\beq
\sum_n\Theta ^{iI_na}_{bI_ni} \equiv \sum_{p<q}S^i_b(p)S^a_i(q).
\eeq

Calculating along the same lines, gives us,
\beqs
\left[\Theta ^{pqr}_{rqp}, \Theta ^{iI_na}_{bI_n i}\right] = \hbar \left(\Theta ^{ijkI_na}_{kjbI_ni} + \Theta ^{iI_njka}_{bI_nikj} - \Theta ^{ijkI_na}_{bjiI_nk} - \Theta ^{iI_najk}_{bI_nkji} \right)
\nonumber \\ \hbar ^3\left(\Theta ^{ijkI_{n-2}a}_{bjiI_{n-2}k} +\Theta ^{iI_{n-2}ajk}_{bI_{n-2}kji}  - \Theta ^{ijkI_{n-2}a}_{kjbI_{n-2}i} - \Theta ^{iI_{n-2}jka}_{bI_{n-2}ikj}\right) \nonumber \\ = \hbar \tilde F^a_b(n) - \hbar ^3\tilde F^a_b(n-2). 
\eeqs
Now, by summing over, $n$ and setting $\hbar =1$ it is clear that most of the terms  from each of the two commutators separately vanish. Now it is not hard to see that $[ \Theta ^{pqr}_{rqp}, \Theta ^{iI_1a}_{bI_1 i}] = O(\hbar)$, i.e $F^a_b(-1)=0$. However,  when one looks at the case $n=0$, one finds that,
\beqs
\left[\Theta ^{pqr}_{rqp},\Theta ^{ai}_{ib}\right] &=& \hbar \left(\Theta ^{aijk}_{jikb} +  \Theta ^{aijk}_{kbji} - \Theta ^{ijak}_{kjib} - \Theta ^{aijk}_{ikjb}\right)\nonumber \\
& &+ \hbar ^2\left(\Theta ^{iaj}_{jib} + \Theta ^{ija}_{jbi} - \Theta ^{iaj}_{bji} - \Theta ^{aij}_{jbi}\right). 
\eeqs 
The term of $O(\hbar)$ cancels with that of  $O(\hbar ^3)$  from the term involving the commutator of the Hamiltonian with $\Theta ^{aI_2i}_{iI_2b}$. However, the term of $O(\hbar ^2)$ remains un-canceled.  Hence, we have, after summing overall $n$ and setting $\hbar = 1$,
\beq
[\Theta ^{pqr}_{rqp}, \sum_{n=0}^{\infty}\Theta ^{aI_ni}_{iI_nb}] =  \left(\Theta ^{iaj}_{jib} + \Theta ^{ija}_{jbi} - \Theta ^{iaj}_{bji} - \Theta ^{aij}_{jbi}\right) \label{viol}
\eeq
Anti-symmetrizing in $a$ and $b$ does not help either, and we have,
\beq
\left[\Theta ^{pqr}_{rqp}, (Q^2)^a_b\right] = \left[\Theta ^{pqr}_{rqp}, \sum_n\left(\Theta ^{aI_ni}_{iI_nb} - \Theta ^{iI_na}_{bI_n i}\right)\right]
= 2\left(\Theta ^{iaj}_{jib} + \Theta ^{ija}_{jbi} - \Theta ^{iaj}_{bji} - \Theta ^{aij}_{jbi}\right) 
\eeq
Before we go on to find the deformed generators that commute with the two loop Hamiltonian, it is interesting to note that, in  (\ref{viol}),  tracing over $a$ and $b$, which is nothing but the term of $O(\frac{1}{u^2})$ in the expansion of the transfer matrix of the one loop dilatation generator,  produces a conserved charge.
\beq
\left[D_4, \sum_{n=0}^{\infty}\Theta ^{jI_ni}_{iI_nj}\right] = \left[D_4, Trt^2\right] =0.
\eeq
So,  the first two terms in the expansion of the trace of the transfer matrix (of the one loop dilatation operator) continue to be conserved at two loops.  

{\bf Yangian charges for the two loop dilatation operator:} After ignoring all the terms proportional to the identity, which are irrelevant for the purposes of exploring conservation laws, the analysis of \cite{beisert-et-al-conformal, long-range-inz} tells us that the dilatation operator in the $SU(2)$ sector has the form,
\beq
D = 2\alpha \Theta ^{ij}_{ji} + 2\alpha ^2\left(\Theta ^{ijk}_{kji} - 4 \Theta ^{ij}_{ji}\right) + O(\alpha ^3), \alpha = \frac{\lambda }{16\pi ^2}.
\eeq  
As mentioned at the beginning of this section, there is no problem with the first Yangian generator.
For the Yangian invariance to be implemented perturbatively, we must look for an expansion of the second generator in the form,
\beq
(Q^2)^a_b = (Q^2_0)^a_b + \alpha (Q^2_1)^a_b + O(\alpha ^2),
\eeq
where $(Q^2_0)^a_b$ is the familiar Yangian generator for the Heisenberg spin  chain,
\beq
(Q^2_0)^a_b = \frac{1}{2}\sum_n\left(\Theta ^{aI_ni}_{iI_nb} - \Theta ^{iI_na}_{bI_n i}\right).
\eeq 
$(Q^2_1)^a_b$ is the new generator to be found. It must satisfy
\beq
\left[\Theta ^{ijk}_{kji}, (Q^2_0)^a_b\right] + \left[\Theta ^{ij}_{ji}, (Q^2_1)^a_b\right] = 0
\eeq 
The first term on the left hand side is known explicitly (\ref{viol}). We can now use a result from the calculations regarding the commutation of the Yangian generators with the one loop Hamiltonian, done in the previous subsection, which is (see the discussion preceeding (\ref{comm-one-loop}))
\beq
\left[\Theta ^{ij}_{ji}, \Theta ^{ai}_{ib} - \Theta ^{ia}_{bi}\right] =  -\left(\Theta ^{iaj}_{jib} + \Theta ^{ija}_{jbi} - \Theta ^{iaj}_{bji} - \Theta ^{aij}_{jbi}\right).
\eeq
Comparing with (\ref{viol}), we see  that we have solved our problem. The Yangian charges that commute with the dilatation operator up to two loops, are,
\beq
(Q^1)^a_b = \Theta ^a_b,
\eeq
\beq
(Q^2_{two-loop})^a_b = \frac{1}{2}\sum_n\left(\Theta ^{aI_ni}_{iI_nb} - \Theta ^{iI_na}_{bI_n i}\right) + \alpha \left(\Theta ^{ai}_{ib} - \Theta ^{ia}_{bi}\right)\label{ytl}.
\eeq

{\bf Relation to the Yangians of long ranged spin chains:}
This deformation of the Yangian charges is consistent with the form of the Yangian generators for long ranged spin chains. In the case of the $su(2)$ invariant spin chains of the Haldane-Shastry type, \cite{long-range-inz, inz-rev, hal-shas, ge-rev} the matrix model representation of the second generator is,
\beq
(Q^2_{long-range})^a_b = \frac{1}{2}\sum_n\frac{1 + t^{|I_n|+1}}{1 - t^{|I_n|+1}}\left(\Theta ^{aI_ni}_{iI_nb} -  \Theta ^{iI_na}_{bI_n i}\right)\label{ylr} 
\eeq
We have employed an analytical continuation of the charges ( in $t$) as is done in the case of the Inozemtsev spin chain\cite{long-range-inz, inz-rev}. $t$ is related to the free parameter $k$ in the Inozemtsev Hamiltonian, which in the limit of large chains takes the form,
\beq
H = \sum_n k^2\left(\sinh ^{-2}(|I_n|+1) +\frac{1}{3}\right)\left( \Theta ^{iI_nj}_{jI_ni}\right)
\eeq
as 
\beq
t = e^{-2k}.
\eeq
If one now identifies, as was done in \cite{long-range-inz} to embed the higher loop corrections to the $SU(2)$ dilatation operator in the Inozemtsev chain,
\beq
\alpha  = \sum_n \frac{1}{4\sinh ^2(kn)},
\eeq
one can see that the expansion of (\ref{ylr}) upto  $O(\alpha )$  exactly reproduces (\ref{ytl}). 
\section{The $SO(n)$ Case:} We now turn to the study of Yangian symmetries present in the full subsector of scalars of the one loop dilatation operator. The matrix model Hamiltonian, (after setting the overall coupling constant to unity, and ignoring the piece proportional to the identity) is,
\beq
H = \Theta ^{ii}_{jj} - 2 \Theta ^{ij}_{ji}.
\eeq
The $R$ matrix for the $SO(n)$ case is\cite{resh-1},
\beq
R(u)_{\mu \nu} = u(g-u)\delta _{\mu \nu}I + (g-u)S_{\nu \mu} +uS_{\mu \nu}, g = (1-\frac{n}{2}). 
\eeq
We are not going to distinguish between the upper and lower indices in this case. What we have above are the matrix elements of the $R$ matrix, thought of as a $n \times n$ matrix. The $S$'s are the usual Weyl operators, satisfying
\beq
[S_{\mu \nu}, S_{\alpha \beta }] = \delta _{\nu \alpha }S_{\mu \beta} - \delta _{\mu \beta}S_{\alpha \nu }.
\eeq
The transfer matrix for a lattice of $n$ sites is,
\beq
T_{\mu \nu}(u) = \tilde{R}(u)^{i_1}_{\mu \nu _1}\tilde{R}(u)^{i_2}_{\nu_2 \nu _3}\cdots \tilde{R}(u)^{i_n}_{\nu _n \nu} = \sum_l \frac{1}{u^l} t^l_{\mu \nu},
\eeq
where,
\beq
\tilde{R}(u)^{i}_{\mu \nu } = \delta _{\mu \nu} I(i) + \frac{1}{u}S_{\nu \mu}(i) + \frac{1}{g-u}S_{\mu \nu }(i),\label{rson}
\eeq
which is nothing but the $R$ matrix divided by a spectral parameter dependent factor, which makes it more conducive to the $\frac{1}{u}$ expansion.
We can identify the first two terms of the expansion of the transfer matrix, which are,
\beqs
t^1_{\mu \nu} = \sum_i L^1_{\mu \nu}(i),  L^1_{\mu \nu}(i) = S_{\nu \mu}(i) -   S_{\mu \nu}(i)\nonumber\\
t^2_{\mu \nu} = \left(\sum_{i<j}L^1_{\mu \alpha }(i)L^1_{\alpha \nu}(j) - g \sum_iL^2_{\mu \nu}(i)\right), L^2_{\mu \nu}(i) = S_{\mu \nu }(i).
\eeqs
Denoting the collective indices $\mu \nu$ as a single $so(n)$ index $a$, we see that the first element of the transfer matrix is Lie algebra valued, and we take it to be the first Yangian generator.
\beq
Q^1_a = t^1_{\mu \nu} = \sum _i L^1_a(i).
\eeq
In the matrix model language,
\beq
Q^1_{\mu \nu} = \Theta ^\nu _\mu - \Theta ^\mu _\nu,
\eeq

and, it is an easy exercise to see that $Q^1$, being the $SO(n)$ generator is conserved. 
Unlike the $SU(n)$ case, the second term in the expansion of the transfer matrix is not Lie algebra valued, and neither is it conserved. 
To get the second non-local charge, we shall have to anti-symmetrize $t^2$ in $\mu $ and $\nu $, to get,
\beq
Q^2_a = t^2_{\mu \nu} - t^2_{\nu \mu} = \frac{1}{2}f_{abc}\sum_{i<j}L^1_b{i}L^1_c(j) + g\sum_iL^1_a(i).
\eeq
Here, the f's are the structure constants of $so(n)$, 
\beq
f_{abc} = f_{\mu \nu, \alpha \beta, \rho \sigma} = \delta _{\mu \rho}\delta _{\nu \alpha} \delta _{\beta \sigma} - \delta _{\mu \alpha}\delta _{\nu \rho }\delta _{\beta \sigma} + \delta _{\nu \beta }\delta _{\alpha \rho} \delta _{\sigma \mu} - \delta _{\mu \beta} \delta _{\rho \alpha} \delta _{\sigma \nu}\label{structure}
\eeq
The matrix model equivalent of the path ordered quantity appearing in the second generator is,
\beq
\sum_{j<k}L^1_{\mu \alpha}(j) L^1_{\alpha \nu}(k) =  \sum_{n=0}^\infty \Theta ^{iI_n\nu}_{\mu I_ni} + \Theta ^{\mu I_n i}_{iI_n\nu} - \Theta ^{\mu I_n\nu}_{iI_ni} - \Theta ^{iI_ni}_{\mu I_n \nu}
\eeq
The second charge is conserved upto boundary terms, i.e its commutator with the Hamiltonian can be written down completely in terms of the $\tilde \Theta $ terms. So it is conserved in $Pl(n)/K(n)$.  The explicit expressions for the boundary terms that result in evaluating the commutator of the the second Yangian generator with the Hamiltonian, on a state of length $n+2, n\geq 0$ is the following,
\beqs
\frac{1}{2}\left[H, Q^2_{\mu \nu}\right] &=& g\left(\tilde \Theta ^{llI_n}_{\mu I_n \nu} - \tilde \Theta ^{\mu I_n \nu}_{I_n ll}\right) - \left(\tilde \Theta ^{l\nu I_n}_{\mu I_n l} - \tilde \Theta ^{lI_n\nu}_{I_n \mu l}\right)\nonumber \\
& &+\left(\tilde \Theta ^{\mu \nu I_n}_{l I_n l} - \tilde \Theta ^{\mu I_n \nu}_{I_n ll} \right) + \left(\tilde \Theta ^{llI_n}_{\mu I_n \nu} - \tilde \Theta ^{lI_nl}_{I_n \mu \nu}\right) -\left(\mu \Leftrightarrow \nu \right)
\eeqs   
Hence we see that the second Yangian generator is conserved in $Pl(n)/K(n)$. That these charges generate the $SO(n)$ Yangian is a straightforward, but laborious exercise. One sees that the following equations,
\beq
[Q^1_a,Q^1_b] = f_{abc}Q^1_c, [Q^1_a,Q^2_b] = f_{abc}Q^2_c,\label{yson1}
\eeq
and the Serre relations,
\beq
\frac{1}{2}f_{a[bc}[Q^2_{d]}, Q^2_a] = \frac{1}{24}f_{bip}f_{cjq}f_{dkr}f_{ijk}\{Q^1_p,Q^1_q,Q^1_r\},\label{yson2}
\eeq
hold.  The $[,]$brackets on the indices on the left hand side of the equation imply the totally antisymmetric combination of the indices enclosed in those brackets. Unlike the $SU(n)$ case, we were unable to any modification of the Yangian generators that would make them conserved on states of any size. However, the above analysis clarifies the relation of the $SO(n)$ Yangian to the transfer matrix of interest to us.  
\subsection{ Continuum Limits and the Semi-Classical Yangians:}
In this final chapter, we shall study the continuum limits of the matrix models described in the previous sections and work out the resultant contraction of the Yangian symmetry.
\subsection{Variational Principles:}  
Ideally, one would like to describe the large $N$ limits of Hamiltonian matrix models as classical dynamical systems. 
This would necessitate  a full description  of the co-adjoint orbit of $Pl(n)$. The co-adjoint orbit would then serve as the phase space of the classical mechanical system.  In the literature on large $N$ limits, such classical mechanical descriptions have been very fruitful, and we shall refer the reader to the review by Yaffe\cite{Yaffe1} for a fuller exposition.  In the absence of direct methods to understand the co-adjoint orbit of the infinite dimensional algebra described above, we shall use the approach of Yaffe to bypass this problem. The basic idea is to pick a standard set of states in the Hilbert space of the matrix model and regard the expectation value of the matrix model operators on those states as the corresponding classical quantities on the phase space. In doing this one must make sure that the set of states is a complete or over-complete one. A typical state, large $N$ state can be characterized by a set of tensors $Z_{i_1\cdots i_{|I|}}$.
\beq
\ket{Z} = Z_{i_1 i_2 \cdots i_{|I|}}\ket{i_1 i_2 \cdots i_{|I|}},
\eeq
These tensors can be regarded as coordinates on the classical phase space. The expectation value of the matrix model operators on such states, i.e 
\beq
\Theta ^I_J \rightarrow \theta ^I_J(Z)
\eeq
where, the classical observable $\theta $ is,
\beq
<Z|\Theta ^I_J|Z> = \sum_lZ^{\star k_1\cdots k_li_1\cdots i_{|I|}k_{l+|I|+1}\cdots k_n}Z_{k_1\cdots k_lj_1\cdots j_{|J|}k_{l+|J|+1}\cdots k_n}.
\eeq
These tensors are taken to be cyclically symmetric, and $\star $ denotes complex conjugation. So far, no real approximation has been made. We have simply translated the quantum mechanical problem in the language of phase space variables. One can now make a factorized ansatz for the tensor $Z$.
\beq
Z_{i_1\cdots i_n} = Z_{i_1}(1)\cdots  Z_{i_n}(n)
\eeq
In making the factorized ansatz, special care needs to be devoted to ensure that whatever flavor ($R$) symmetry is present in the matrix model Hamiltonian be appropriately captured by  the ansatz. This point has already been elaborated upon in \cite{tseytlin-1, tseytlin-2}. The $Z$'s are clearly coordinates in $CP^{n-1} $, where $n$ is the range of the bits. So, it makes sense to impose the norm $Z^{\star i}(l)Z_i(l) =1$. This is all that one needs in the  $SU(n)$ case. However, in the case of $SO(n)$, one needs to impose the added constraint $Z^{\star i}(l)Z^{\star i}(l) =  Z_i(l)Z_i(l) =0$. This further constraint has an interpretation as a local BPS condition\cite{tseytlin-1, tseytlin-2}.The classical observables corresponding to the factorized ansatz take on the following forms,
\beqs
 <Z|\Theta ^I_J|Z> =  \sum _a (Z^\star (a)^{i_1}Z(a)_{j_i}Z^\star (a+1)^{i_2}Z(a+2)_{j_2}\cdots \nonumber \\Z^\star (a + |I| -1)^{i_{|I|}}Z(a + |J| -1)_{j_{|J|}})
\eeqs
One can define classical spin variables as,
\beq
(Z^\star )^j(a)Z_i(a) = S^j_i(a), 
\eeq
which satisfy the Poisson brackets inherited from the underlying $CP^{n-1}$ manifold. 
\beq
\{ S^j_i(a),S^k_l(b)\} = \delta _{ab} \left(\delta ^k_i  S^j_l(a) - \delta ^j_lS^k_i(a) \right) \label{spin-poisn}
\eeq
So, the classical matrix model variables go over to Hamiltonians of classical spin chains.
\beq
\theta ^I_J(Z) \equiv \theta ^I_J(S) = \sum _a S^{i_1}_{j_1}(a)S^{i_2}_{j_2}(a+1)\cdots S^{i_{|I|}}_{j_{|J|}}(a + |I| -1),
\eeq
where $|I| = |J|$ is implied. 
Since the underlying $Z$'s belong to the complex projective space, the spin matrices can be thought of as the gauge invariant coordinates on $CP^{n-1}$. It is also consistent to require $S^2 =S$.
\subsection{SU(n) Sector:}For the one loop SU(n) invariant dilatation operator, the factorized ansatz produces the classical Hamiltonian,
\beq
H_{su(n)} = \Theta ^{ij}_{ji} \equiv \lambda \sum_lS^i_j(l)S^j_i(l+1),
\eeq
with the Poisson brackets given in (\ref{spin-poisn}). We have absorbed various factors of $\frac{1}{16\pi ^2}$  by  rescaling the spin matrices. The continuum limit is now easy to take. Denoting the total length of the state by $J$ and holding $\tilde \lambda = \frac{\lambda }{J^2}$ fixed, one gets,
\beq
H_{su(n)} = J\tilde \lambda \int dx Tr\left(S\partial ^2_xS\right).
\eeq
One can define the matrix,
\beq
M^i_j = 2S^i_j(x) - \delta ^i_j = 2Z^{\star i}(x)Z_j(x) - \delta ^i_j: M^2 =1,\label{sphere}
\eeq
so the continuum Hamiltonian now reads,
\beq
H_{su(n)}= J\frac{\tilde \lambda}{4}\int dx Tr\left(M\partial ^2_xM\right), M^2 =1
\eeq
which is the $SU(n)$ generalization of the classical Heisenberg model. This construction works for any 
$SU(n)$.  This particular sigma model is known to be integrable, and the underlying symmetry is the semiclassical $SU(n)$ Yangian. For a detailed account of the $SU(2)$ case see \cite{fadbook, bernard-babelon}. For the sake of completeness let us work out the basic ideas leading to its integrability. The equations of motion of the sigma model,
\beq
\partial _t M = \partial[M,\partial _x M]
\eeq 
can be interpreted as the condition for the conservation of a $SU(n)$ valued current. We have divided out the Hamiltonian by $J$ and set $\tilde \lambda =1$ in writing down the equations of motion. This motivates the introduction of the following Lax connection for the problem.
\beqs
A_x(\lambda) =  \frac{1}{\lambda }M, \nonumber \\
A_t(\lambda ) = \frac{1}{\lambda }[M,\partial _x M] -\frac{2}{\lambda ^2}M,
\eeqs
where $\lambda $ is the spectral parameter. The equations of motion can easily be seen to be equivalent  to the condition for this connection to be flat,
\beq
[\partial _x + A_x, \partial _t + A_t]=0
\eeq
This naturally leads to the monodromy matrix, 
\beq
T(x,\lambda ) = P\exp \left(-\int_0 ^x A_x(\lambda, y )dy\right).
\eeq
The Poisson bracket relations satisfied by the monodromy matrix can be written down as a set of classical Yang-Baxter relations,
\beq
\{T(\lambda )\stackrel{\otimes}{,}T(\mu)\} = [r(\lambda - \mu), T(\lambda )\otimes T(\mu)],
\eeq 
where the classical 'r' matrix \cite{fadbook} is,
\beq
r(\lambda) = \frac{2}{\lambda }P.
\eeq
$P$ is the permutation operator on $V \otimes V$. The factor of two in front of the permutation operator has to do with the particular parameterization (\ref{sphere}) of $CP^{n-1}$ that we employed. \footnote{The in the convension used above, the operators on the tensor product of the auxilliary vector space $V$ with itself, take on the following form in components. Let $A$ and $B$, be two $n \otimes n$ matrices, then,  
\beq
\left(A \otimes B\right)^{ik}_{jl} = A^i_jB^k_l.
\eeq
}
 As in the case of the spin systems,  the transfer matrix can also be regarded as the generator of the Yangian charges, except, now that we are dealing with a classical theory, the Yangian symmetry is implemented by a Lie-Poisson action, and it is no longer a quantum group symmetry. For a very useful discussion of the Lie-Poisson symmetry of the $SU(2)$ cousin of this model, we shall refer to \cite{bernard-babelon}, and for a general exposition on Yangian symmetries in  integrable models see\cite{clss-yang, leclair-yangian, mckay-yangian-2}. To be able to write down the semi-classical Yangian generators in a compact fashion, we shall, as we did before, denote the upper and lower $SU(n)$ indices compactly by a single Greek index  e.g, $M^\alpha \equiv M^a_b$, and the $SU(n)$ structure constants become,
\beq
f^{\alpha \beta \gamma } = f^{acp}_{bdq} = \delta ^c_b\delta ^a_q\delta ^p_d - \delta ^a_d \delta ^c_q \delta ^p _b
\eeq
The Yangian generators can be read off from the transfer matrix, they are,
\beqs
(Q^1)^\alpha  = \int _0^J M^\alpha (x),\nonumber \\
(Q^2)^\alpha = \frac{1}{2}f^{\alpha \beta \gamma }\int_0^J M^\beta (x)dx\int_0^xM^\gamma (y)dy.
\eeqs
These are the only independent generators, and the entire transfer matrix can be generated by evaluating iterated Poisson brackets of these generators with themselves. They satisfy the relations of the semi-classical $SU(n)$ Yangian, which are,

\beqs
\{(Q^1)^\alpha, (Q^1)^\beta \} = f_{\alpha \beta \gamma}(Q^1)^\gamma \nonumber \\
\{(Q^1)^\alpha, (Q^2)^\beta \} = f_{\alpha \beta \gamma}(Q^2)^\gamma 
\eeqs
and the  semi-classical Serre relations,
\beq
\frac{1}{2}f^{\alpha [\beta \gamma}\{(Q^2)^{\delta ]}, (Q^2)^\alpha \} = \frac{1}{4}f^{\beta \epsilon \sigma }f^{\gamma \kappa \mu}f^{\delta \tau \nu }f^{\epsilon \kappa \tau}\left((Q^1)^\sigma (Q^1)^\mu (Q^1)^\nu \right),
\eeq
This is the residual Yangian symmetry that survives the continuum limit of the matrix models, and manifests itself as the underlying symmetry of the $SU(n)$ reduction of the string sigma model. This analysis is, in a sense, complimentary to the one presented in \cite{minahan-et-al-su2}, where, the relation of the transfer matrix of the $SU(2)$ spin chain to its continuum limit was utilized to present a unified approach towards  understanding the various results and integrable structures found in this particular sector. Since the Yangians are also the generators of the transfer matrices, the contraction of the Yangian symmetries implies that at the semi-classical level, one should replace the 'quantum'spins of the spin chains by the corresponding classical quantities. As was pointed out, in \cite{minahan-et-al-su2}, this quantum to classical correspondence is difficult to motivate at the level of the transfer matrices, however, it appears to be quite natural from the point of view of  variational principles. Alternatively, the classical nature of the continuum limit can now be understood as a consequence of the contraction of the Yangian symmetry of the dilatation operator. 
\subsection{SO(n) Sector:}
Keeping the above  discussion in mind, we can now proceed to study the Yangian symmetries of the continuum limit of the full $SO(n =6)$ sector of scalars. 
In this sector, the  matrix model corresponding to the one-loop Hamiltonian is,
\beq
\Gamma = \frac{\lambda }{16\pi ^2}:Tr\left(a^{\dagger i}a^{\dagger i}a_ja_j - a^{\dagger i}a_ja^{\dagger i}a_j +2(a^{\dagger i}a^{\dagger j}a_ja_i - a^{\dagger i}a^{\dagger j}a_ia_j)\right):,\label{dilop}
\eeq
Although $SO(6)$ is the case of interest to us, we shall do the following analysis for $SO(n)$ in general. 
As in the $SU(n)$ case, we are going to take a factorized variational ansatz,
\beq
|Z> = \frac{1}{\sqrt{N^n}}Z_{i_1}(l)Z_{i_1}(l+1)\cdots Z_{i_n}(l+(n-1))Tra^{\dagger i_1}\cdots a^{\dagger i_n}|0>
\eeq
As it stands, the variational ansatz has $SU(n)$ instead of $SO(n)$ invariance. But, as was pointed out in\cite{tseytlin-1, tseytlin-2}, see also\cite{mikhailov-null},  we can impose  $SO(n)$ invariance by requiring the vectors to satisfy the local BPS condition,
\beq
Z^*_iZ^*_i = Z_iZ_i =0 \mbox{as well as the requirement} Z^*_iZ_i = 1 \label{cond1}
\eeq
The classical Hamiltonian that one gets from this is,
\beqs
H_{so(n)} = <V|\Gamma |V> &=& \frac{\lambda }{16\pi ^2}\sum_l  ( Z^* _i(l)Z_j(l)Z^* _i(l+1)Z_j(l+1) \nonumber \\ & &- 2 Z^* _i(l)Z_j(l)Z^* _j(l+1)Z_i(l+1)) 
\eeqs
One may now proceed to take the continuum limit, which generates for us a variant of the $CP^{n-1}$ model. Once again, after absorbing various factors of $\frac{\lambda }{16\pi ^2}$ in the redefinition of the $Z$,s, the continuum action reads,
\beq
H_{so(n)} = J\frac{\tilde \lambda }{4} \int dx \left(|\partial Z|^2 - |Z^* \partial Z|^2\right) -\mu_1(Z^*. Z -1)-(\mu_2 Z. Z + cc).\label{hsonz}
\eeq
$\mu_1 , \mu_2$ are the Lagrange multipliers that enforce the constraints described above. Comparison of particular solutions of this continuum Hamiltonian to the Bethe ansatz solutions obtained by Minahan and Zarembo was carried out in \cite{tseytlin-2,tseytlin-1}. See also, \cite{twisted-sectors}. 
To make the $SO(n)$ invariance manifest, one can now introduce the antisymmetric matrix,
\beq
m_{ij} = Z_iZ^*_j - Z^*_iZ_j,
\eeq
which enjoys the following properties,
\beq
Tr m^2 =2, m^3 = m, m(\partial m) m = 0 \nonumber \label{prop1}
\eeq
\beq 
m^2 (\partial m) + (\partial m)m^2 = (\partial m) = (\partial m^3), \nonumber   [m , [m , \partial m ]] = \partial m.\nonumber
\eeq
These properties are consequences of the relations (\ref{cond1}) satisfied by the $Z$'s. The Poisson brackets satisfied by $m$ is,
\beq
\{m_{ij}(x), m_{kl}(y) \} = \delta (x-y) \left(\delta _{jk}m_{il}(x) + \delta _{il}m_{jk}(x) - \delta _{ik}m_{jl}(x) - \delta _{jl}m_{ik}(x)\right)
\eeq
In terms of this matrix, the Hamiltonian is,
\beq
H_{so(n)} = \lambda ' \int Tr\left(\partial m \partial m\right), \lambda ' = J\frac{\tilde \lambda}{4}\label{hsonm}
\eeq
In the form sigma model Hamiltonian reported in \cite{tseytlin-2}, there is an additional term proportional to $\int Tr(m\partial m)^2$. We note, from the identities reported above that, such a  term is identically zero. 
 
The equations of motion resulting from this Hamiltonian are very much like those that follow from the matrix $SU(n)$ Hamiltonian,
\beq
\frac{\partial m}{\partial t} = \alpha \partial [m, \partial m], \alpha = -4 \lambda '
\eeq
Following the methods employed in the case of the Heisenberg model, one can introduce a Lax connection,
\beqs
A_x = \frac{1}{\lambda } m \nonumber \\
A_t = \frac{1}{\lambda } \alpha [m , \partial m] - \frac{\alpha }{\lambda ^2} m \label{laxson}
\eeqs
The using (\ref{prop1}), one can see that the equations of motion are equivalent to the flatness condition,
\beq
\left[ \partial _t + A_t, \partial _x + A_x \right] = 0.
\eeq
To be able to apply Bethe anzatz techniques to this sigma model, and to compare it with its quantum counterpart, it is important to be able to write down the commutation relations between the components of Lax connection as classical Yang-Baxter equations. 
 
The Poisson brackets between the Lax connection may now be expressed using the classical $r$ matrix,
\beq
\{A_x(x,\lambda ) \stackrel{\otimes }{,} A_x(y, \mu) \} = \delta (x-y) [r(\mu - \lambda) , A_x((x,\lambda )\otimes I +  I \otimes  A_x(y, \mu)], \label{fpb-son}
\eeq
where,
\beq
r(\lambda ) = \frac{1}{\lambda } \left(K - P\right).
\eeq
$K$ and $P$ are the usual trace and permutation operators on $V\otimes V$, i.e. $K (V \otimes W) = V.W e^k \otimes e^k$, and $P(V \otimes W) = W \otimes V$, where $e^k$'s are the basis vectors in $V$. It is also understood, that, for any two $n\times n$ matrices $A$ and $B$, $(A \otimes B)_{ij,kl} = A_{ik}B_{jl}$.   
To see that (\ref{fpb-son}) is indeed true, one first notices that
\beq
\{m(x) \stackrel{\otimes }{,} m(y)\} = \delta (x-y)[K-P, m(x) \otimes I] \label{step-1}.
\eeq
Furthermore, the following  set of equations are true,
\beq
[K , m(x) \otimes I] = \left(\frac{1}{\lambda }- \frac{1}{\mu}\right)[K, A_x(x, \lambda ) \otimes I + I \otimes A_x(x,\mu)]\label{step-2},
\eeq
and
\beq
[P , m(x) \otimes I] = \left(\frac{1}{\lambda }- \frac{1}{\mu}\right)[P, A_x(x, \lambda ) \otimes I + I \otimes A_x(x,\mu)]\label{step-3}.
\eeq
In deriving these equations, we have used the defining equation (\ref{laxson}) for $A_x$, along with the identities, 
\beq
K(m \otimes I) = -K(I \otimes m), (m \otimes I)K = - (m \otimes I)K,
\eeq
which follows from the antisymmetry of $m$, and the standard relation 
\beq
[P,(A\otimes B)] = -[P,(B\otimes A)], 
\eeq
for any  two $n \times n$ matrices $A$ and $B$. The Poisson bracket for the Lax connection (\ref{fpb-son}), now follows from the equations (\ref{step-1}$\cdots$  \ref{step-3}). 
The generators one may now construct  the monodromy matrix,
\beq
T(\lambda ) = P exp\left(\int _0 ^J A_x(y)dy\right).
\eeq
Following standard techniques \cite{fadbook}, one can see that (\ref{fpb-son}) imply that the  Poisson brackets between the matrix elements of the monodromy matrix, for different values of the spectral parameter are,
\beq
\{T(\lambda ) \stackrel{\otimes }{,}  T(\mu )\} = [r(\lambda - \mu), T(\lambda ) \otimes  T(\mu )]. 
\eeq
As usual, realization of the Poisson brackets through Lie brackets, as above, implies that $Tr T(\lambda) $ is the generating function for an infinite family of conserved charges, which are in involution with each other.  

The transfer matrix can now be seen to generate the semiclassical $SO(n)$ Yangian. As before, (see for example the discussion preceeding (\ref{structure})), we can label the matrix indices, by a single index $a$. The Lie algebra valued Yangian charges that one gets from the expansion of the transfer matrix are,
\beqs
Q^1_a = \int _0^J m_a(x)dx \nonumber \\
Q^2_a = \frac{1}{2}f_{abc}\int _0^J m_{b}(x)dx\int _0^xm_c(y)dy.
\eeqs
The Poisson bracket relations of these charges now generate the semi-classical $SO(n)$ Yangian. One gets,
\beqs
\{Q^1_a, Q^1_b\} = f_{abc}Q^1_c \nonumber \\
\{Q^1_a, Q^2_b\} = f_{abc}Q^2_c,
\eeqs
as well as the semi-classical Serre relations,
\beq
\frac{1}{2}f_{a[bc}\{Q^2_{d]}, Q^2_a\} = \frac{1}{4}f_{bip}f_{cjq}f_{dkr}f_{ijk}\left(Q^1_pQ^1_qQ^1_r\right),
\eeq
where, as before, the $[,]$brackets on the indices on the left hand side of the equation imply the totally antisymmetric combination of the indices. This is the underlying Lie-Poisson symmetry of the semi-classical limit. 

{\bf Comments on relations between integrable structures:} Since, (\ref{hsonz}) and (\ref{hsonm}) are two ways of looking at the same classical field theory, it is now quite clear that the part of the full Yangian symmetry of the classical string theory in $AdS_5 \times S^5$ which is retained in its  $SO(n)$ reductions  \cite{tseytlin-1} \cite{tseytlin-2} is the semi-classical $SO(n)$ Yangian. Moreover, the symmetry of the reduced sigma model being the semi-classical counterpart of the  one underlying the dilatation operator generates for us a direct relation between the integrable structures present on the gauge and string theory sides. This is quite like the one we had in the $SU(n)$ case. As before, the classical-quantum correspondence  between the Yangian symmetries implies that the  $r$ and monodromy matrices, of the sigma model are nothing but the classical limits of the corresponding matrices appearing in the Bethe ansatz for the Minahan-Zrembo spin chain. Although this is implicit in the analysis done up to this point, there is a particularly transparent way to see this. As is usually the case in quantum inverse scattering theory, the quantum  'R' matrix has a natural deformation parameter in it. At the risk of some abuse of notation, let us call this $\hbar $ as well. If one chooses to display this explicitly, then the formula for the $'R'$ matrix for  $SO(n)$ invariant spin chains(\ref{rson}) takes the form \cite{pasq},
\beq
\tilde R(\lambda ) = I + \frac{\hbar }{\lambda} P + \frac{1}{\hbar g -\lambda}K  = I -\hbar r (\lambda ) +O(\hbar ^2),
\eeq
where $r(\lambda )$ for the $SO(n)$ case was derived in (\ref{fpb-son}).
This relation makes it very clear that the $'r'$ matrix for the sigma model is the classical limit of the one describing the spin chain. Moreover, because of the fact that in the $SO(n)$ case, the $R$ matrix is the Lax connection, the same classical-quantum correspondence is true for the monodromy matrices of the two theories as well. This completes for us a map between the fundamental integrable structures which are responsible for the integrability of the one loop (bosonic) dilatation operator and its corresponding continuum limit. In the light of the identification of the continuum limit of the dilatation operator and the semiclassical string actions\cite{tseytlin-1, tseytlin-2}, this analysis also relates the integrable structures on the gauge theory and gravity sides, in the  $SO(n)$ sector. 

The existence of the Zero curvature condition, and the classical $r$ matrix for the sigma model, implies that one should be able to carry out the complete integration of the problem, i.e. it should be possible, like in the case of the classical Heisenberg field theory \cite{fadbook}, to accomplish an explicit change of variables to the action-angle coordinates of this theory. Moreover, the relation between the integrable structures presented above should allow one to find the semi-classical counterparts of the Bethe eigenstates found in \cite{minahan-spin}, putting the analysis of the $SO(n)$ sector on the same footing as the $SU(2)$ one \cite{minahan-et-al-su2}. We hope to report on this in the near future.

{\bf Lie-Poisson actions:} Before concluding the discussion on the $SO(n)$ sigma model, let us examine how the semi-classical Yangian symmetry acts on the basic degrees of freedom of the theory.  From the form of the Hamiltonian, it is obvious that any $SO(n)$ transformation is a symmetry of the Hamiltonian, i.e the infinitesimal transformation,
\beq
\delta ^0 m = [m,m_0],
\eeq
where, $m_0$ is a constant $SO(n)$ matrix leaves the Hamiltonian invariant. However the semi-classical Yangian invariance implies that,
\beq
\delta ^n m(x) = [m(x),m_n(x)], \mbox{where}, \partial _x m_n(x) = [m,m_{n-1}(x)]
\eeq
are all symmetries of the Hamiltonian. These symmetries are a consequence of the Lie-Poisson symmetries generated by the transfer matrix. Indeed, for any constant $SO(n)$ matrix, $m_0$, one can define the generator of these symmetries, as,
\beq
\left(Tm_0T^{-1}\right)(x,\lambda ) = \sum_{n}\lambda ^{-k}m_k(x),
\eeq
and the equations satisfied by $m_n$ are a consequence of the ones satisfied by $T$.  Clearly, one can now proceed to express $m_n$ in terms of the Yangian charges, (the above equation is a prescription for doing that), and express the symmetry transformations, as Poisson brackets of the charges with the basic dynamical variable $m(x)$.  The answer can be summed up succinctly in the following $SO(n)$ generalization of the formula for Lie-Poisson transformations of the $SU(2)$ Heisenberg model \cite{bernard-babelon}.
\beq
\delta ^nm(x) = \oint\frac{d\lambda }{2\pi i}\lambda ^n tr_1\left((m_oT^{-1}(\lambda)\otimes 1)\{T(\lambda \otimes 1, 1\otimes m(x)\}\right)
\eeq

\subsection{Concluding Remarks and Future Directions:} So far we have seen that Yangian symmetries play a natural role, at least in low orders in perturbation theory, in forming a bridge between the integrable structures present in SYM and its gravity dual. At the same time, we can also see, that the dilatation operator and the continuum sigma models can be described quite efficiently using Hamiltonian matrix models, which are also useful in describing  these Hopf algebraic symmetries and their semi-classical counterparts. Clearly, a lot of questions and possibilities are yet to be explored. A natural question is whether or not the matrix model provides a direct way of diagonalizing the dilatation operator without referring to the spin chains. The answer to that, at least in principle, is in the affirmative. For example, in the examples discussed previously, the Yangian generators, as well as the Hamiltonians,  have a faithful realizations completely within $Pl(n)$. Once, one obtains such a realization, then one can  proceed to solve the problem of diagonalizing the Hamiltonian using the  representation theory of the Yangian generators. This, for example, was the approach which was used quite successfully in the context of the Haldane-Shastry spin chains; see \cite{haldane-yan} for a review. The employment of such symmetry considerations in diagonalizing the integrable Hamiltonians   would make the approach self-sufficient. Hence, it is necessary  to work out a few examples of such solutions for the matrix models describing various sectors of the dilatation operator of superconformal Yang-Mills theory. We hope to report on this issue in the near future\cite{aa-inprep}. One can hope that understanding integrability of matrix models based on such representation theoretic grounds would probably improve our understanding the 'dynamical' spin chains which arise in higher orders in perturbation theory. In a sense, such a diagonalization procedure would be to a large extent independent of the particular form of the Hamiltonian. In the language of the algebraic Bethe ansatz, this is akin to knowing the Bethe equations without knowing in  detail  the underlying Hamiltonian. Such themes have appeared in various places in the context of understanding integrability within the AdS-CFT correspondence.  For a recent example of such a circumstance, we shall refer the reader to the recent work presented in \cite{frolov-et-al-baqs}, where the Bethe equations for the diagonalization of the $SU(2)$ sector of the sigma model have been proposed. It should be noted however, that Matrix model techniques can also be employed to capture the more conventional algebraic Bethe ansatz techniques employed in the literature on spin chains. This has been elaborated on in the appendix.

On a slightly different note, it is possibly worth while to understand what kind of quantum group symmetries lie behind the pp-wave limit of matrix theory, where too, integrability to quite high orders in perturbation theory has been shown to exist\cite{plefka-matrix-1,plefka-matrix-3,plefka-matrix-2}. 

Apart from these issues, a precise of understanding the quantum group symmetries beyond the $SU(2)$ sector of the dilatation operator remains a reasonably  open issue. The sigma model realization of the continuum limits of the higher loop dilatation operators and  their relation to the Yangian symmetries of the classical string sigma model too requires further analysis.  

\section{Appendix A:}
{\bf Finite Rank Operators and The Ideal:} As mentioned in the introduction, the planar Lie algebra $PL(n)$, where $n$ stands for the number of matrices involved, or alternatively, the range of the string bit indices, has an ideal $K(n)$, generated by the $\tilde \Theta $ operators.  The elements of the ideal are can be written as finite linear combinations of the elements of the Lie algebra.
\beq
\tilde \Theta ^I_J = \tilde \Theta ^I_J - \tilde \Theta ^{Ii}_{Ji} =  \tilde \Theta ^I_J - \tilde \Theta ^{iI}_{iJ}. \label{tildedef}
\eeq
The implications of the two ways expressing the same elements of the ideal have been commented on earlier (see section(2.1).

The elements of $K(n)$ act as Weyl operators on the states of the matrix model. They form a Lie algebra by themselves,
\beq
[\tilde \Theta ^I_J, \tilde \Theta ^K_L] = \delta ^K_J\tilde \Theta ^I_L - \delta ^I_L\tilde \Theta ^K_J. 
\eeq
These operators form an ideal, as their commutation relations with the elements of $PL(n)$ are of the following form,
\beq
[\Theta ^I_J, \tilde \Theta ^K_L] = \delta ^K_J\tilde \Theta ^I_L + \sum_{K=K_1K_2} \delta ^{K_1}_J\tilde \Theta ^{IK_2}_{L} - \delta ^I_L\tilde \Theta ^K_J - \sum_{L=L_1L_2} \delta ^{I}_{L_2}\tilde \Theta ^{K}_{L_1J},
\eeq
i.e. the commutator of elements of $K(n)$ with those of $Pl(n)$ is a finite linear combination of the elements of $K(n)$. However, one should be careful, and note that, although finite linear combinations of the elements of $K(n)$ can be regarded as finite rank operators, infinite linear combinations of these operators carry the same information as the full $Pl(n)$. For example, the relation ($\ref{tildedef}$) can be inverted iteratively, to yield,
\beq
\Theta ^I_J = \sum_K\tilde \Theta ^{IK}_{JK}, 
\eeq
i.e, elements of $Pl(n)$ can be regarded as infinite linear combinations of the elements of $K(n)$. However, when one deals with states of a fixed sizes, only a finite number of terms in this infinite sum survive. Now, in the $SU(2)$ sector, the length of the state is always conserved, or in the language of spin chains, the chains are not dynamical. So by fixing the length of the state, $J$, one can recast everything in terms of the elements of $K(n)$. The advantage of doing this is that it allows one to map spin chain operators, which are not necessarily translationally invariant in the matrix model language, thus enabling us to construct the Lax matrix in terms of the elements of $Pl(n)$. For example, let us look at the operator, $S^i_j(2)$, the spin operator acting on the second lattice site, on a chain of length $J$. This can be represented by,
\beq
S^i_j(2) = \tilde \Theta ^{liK}_{ljK}, |K| = J-2    
\eeq
It should be kept in mind that  we are tacitly ignoring the cyclic symmetry of the matrix model states in carrying out this identification. 
More generally, if we have some operator $O$, not necessarily translation invariant, acting on the a chain of length $J$, we can form out of that operator, the corresponding finite rank matrix model operator,
\beq
\tilde \Theta (O) = \sum_{|I| = |K| = J}\Psi ^I_K(O)\tilde \Theta ^I_K,
\eeq 
where, the tensor $\Psi ^I_K(O) = <K|O|I>$, is the matrix element of the spin chain operator on two states of the spin chain. It is now straight forward to see that,
\beq
\tilde \Theta (O_1) \tilde \Theta (O_2) = \tilde \Theta (O_1O_2),
\eeq
and,
\beq
[\tilde \Theta (O_1) \tilde \Theta (O_2) , \tilde \Theta (O_2) \tilde \Theta (O_1)] = \tilde \Theta ([O_1, O_2]).\label{rep}
\eeq
Thus, in this language, it is very clear that elements of $Pl(n)$ carry a representation of the associative algebra of spin operators. The matrix elements of the Lax matrix, which as we recall are spin operators acting on specific sites (say the $n$th one), is,
\beq
L^i_j(u)_n = \frac{1}{J}(u\tilde \Theta ^I_I \delta ^i_j+ i\sum_{\stackrel{I=I_1aI_2}{|I_1|=n-1, |I|=J}}\tilde \Theta ^{I_1iI_2}_{I_1jI_2}) = \tilde \Theta (I_n\delta ^i_j + iuS^i_j(n)) 
\eeq
Now, one may generate the transfer matrix, for a state of length $J$, which is simply, $\tilde \Theta_J (T^i_j(u))$. 
\beq
\tilde \Theta_J (T^i_j(u)) = \frac{1}{J}\tilde \Theta (I_1\delta ^i_{a_1} +i uS^i_{a_1}(1))\tilde \Theta (I_2\delta ^{a_1}_{a_2} +i uS^{a_1}_{a_2}(2))\cdots \tilde \Theta (I_1\delta ^{a_{J-1}}_{j} +i uS^{a_{J-1}}_{j}(J))
\eeq
The above matrix model realization of the transfer matrix satisfies the Yang-Baxter equations. Moreover, by summing over the variable $J$ and by using the definition of the $\tilde \Theta $ operators,  we recover the simple forms that we had earlier, in terms of $\Theta $,
\beq
\sum_J \tilde \Theta_J (T^i_j(u)) = I + \frac{1}{u}\Theta ^i_j + \frac{1}{u^2}\sum_I\Theta ^{iIa}_{aIj} + \cdots
\eeq
The point of carrying out this exercise makes two things clear. Firstly we see from (\ref{rep}) in rather explicit form that commuting quantities in the spin chain description lead to commuting matrix model operators. Secondly, this generates a systematic way formulating the integrability of the quantum mechanical matrix model in terms of $R$ and $T$ matrices. It should however be noted that this notion of integrability of the matrix model is probably the beginning of a more general understanding of the issue. 

{\bf Acknowledgments:} We are happy to thank  Massimo Bianchi, Ashok Das, Sumit Das, Sergey Frolov, Rajesh Gopakumar, Arsen Melikyan and Matsuo Sato for various useful discussions. We are also  thankful  to Massimo Bianchi, Sumit Das, Sergey Frolov and Rajesh Gopakumar for their comments on an earlier version of the manuscript, and to Sergey Frolov for a correspondence about Yangian symmetries at higher loops. This work was supported in part by US Department of Energy grant number DE-FG02-91ER40685.

\bibliography{abhishekbib}

\end{document}